\documentclass[aps,amssymb,amsmath,amsfonts,pra,twocolumn,superscriptaddress,floatfix,longbibliography]{revtex4-1}
\usepackage{graphicx}
\usepackage[caption=false]{subfig}
\usepackage{soul}

\usepackage{multirow}
\usepackage{xcolor}

\usepackage[colorlinks=true, urlcolor=blue, citecolor=blue,linkcolor=blue,citebordercolor={1 0 0},linkbordercolor={0 0 1}]{hyperref}

\begin{document}

\title{The Meta-Variational Quantum Eigensolver (Meta-VQE): Learning energy profiles of parameterized Hamiltonians for quantum simulation}
\author{Alba Cervera-Lierta}
\affiliation{Chemical Physics Theory Group, Department of Chemistry, University of Toronto, Canada.}
\affiliation{Department of Computer Science, University of Toronto, Canada.}
\author{Jakob S. Kottmann}
\affiliation{Chemical Physics Theory Group, Department of Chemistry, University of Toronto, Canada.}
\affiliation{Department of Computer Science, University of Toronto, Canada.}
\author{Al\'an Aspuru-Guzik}
\affiliation{Chemical Physics Theory Group, Department of Chemistry, University of Toronto, Canada.}
\affiliation{Department of Computer Science, University of Toronto, Canada.}
\affiliation{Vector Institute for Artificial Intelligence, Toronto, Canada.}
\affiliation{Canadian  Institute  for  Advanced  Research  (CIFAR)  Lebovic  Fellow,  Toronto,  Canada}

\begin{abstract} 
We present the \emph{meta-VQE}, an algorithm capable to learn the ground state energy profile of a parametrized Hamiltonian. By training the meta-VQE with a few data points, it delivers an initial circuit parametrization that can be used to compute the ground state energy of any parametrization of the Hamiltonian within a certain trust region. We test this algorithm with a XXZ spin chain, an electronic H$_{4}$ Hamiltonian and a single-transmon quantum simulation. In all cases, the meta-VQE is able to learn the shape of the energy functional and, in some cases, resulted in improved accuracy in comparison to individual VQE optimization. The meta-VQE algorithm introduces both a gain in efficiency for parametrized Hamiltonians, in terms of the number of optimizations, and a good starting point for the quantum circuit parameters for individual optimizations. The proposed algorithm can be readily mixed with other improvements in the field of variational algorithms to shorten the distance between the current state-of-the-art and applications with quantum advantage.
\end{abstract}

\maketitle

\section{Introduction}

Variational quantum algorithms (VQAs) are one of the key tools for the era of noisy intermediate-scale quantum (NISQ) computation \cite{preskill_quantum_2018,NISQreview} and beyond, due to their natural method of optimization. Their hybrid quantum-classical structure exploits the current advantages of both worlds: a quantum circuit is used to compute the expected values of some observable, whereas a classical subroutine finds the optimal parameterization of this quantum circuit. The continuous parameterization of the quantum gates allows us to adjust their arguments to partially compensate for the effect of noisy qubits and imperfect operations. This approach has opened up the possibility of finding applications of quantum computing in the near term, without the need for quantum error correction.

A  VQA  can  be  divided  into  three  principal  blocks:  the preparation of the initial state, the preparation of a parameterized quantum circuit, and the measurement and construction of the cost function. The first step is crucial for starting the algorithm in the Hilbert-space region where the solution is likely to be. The second block guides the algorithm around a particular region of the space of quantum states, and therefore a circuit with high expressibility [3] or a physically inspired ansatz is required. The last block computes the expected value of some operator with the final state of the circuit and constructs a cost function that is then minimized with a classical subroutine. The classical minimizer proposes a new set of values for the quantum circuit block, repeating the loop until convergence or the desired precision is achieved.

The first VQA proposed was the variational quantum eigensolver (VQE) \cite{peruzzo_variational_2014, mcclean_theory_2016}. This algorithm, originally proposed for molecular systems, tries to find the ground-state energy of a given Hamiltonian by variationally minimizing its expectation value with a parameterized quantum circuit. The cost function of this algorithm is the expected value of the model Hamiltonian. The variational principle states that this value is an upper bound on the ground-state energy, and so everything reduces to minimizing this value by fine-tuning the parameters of the circuit. There are different methods to design both the initial state and the quantum circuit ansatz when the model is a molecule \cite{mcclean_compact_2015,Romero2018,cao_quantum_2019, yordanov_efficient_2020}, an integrable condensed-matter model \cite{kraus_compressed_2011,hebenstreit_compressed_2017, schmoll_kitaev_2017,verstraete_quantum_2009,cervera-lierta_exact_2018,montanaro_compressed_2020}, or a more general computational problem \cite{QAOA}. However, there is not a general approach for
other kinds of Hamiltonian. Even in quantum chemistry, where good heuristics for circuit construction and initialization are known, layerwise extension of the circuits can lead to similar initialization problems \cite{lee2018generalized}.

\begin{figure*}[t]
\centering
\includegraphics[width=1.2\columnwidth]{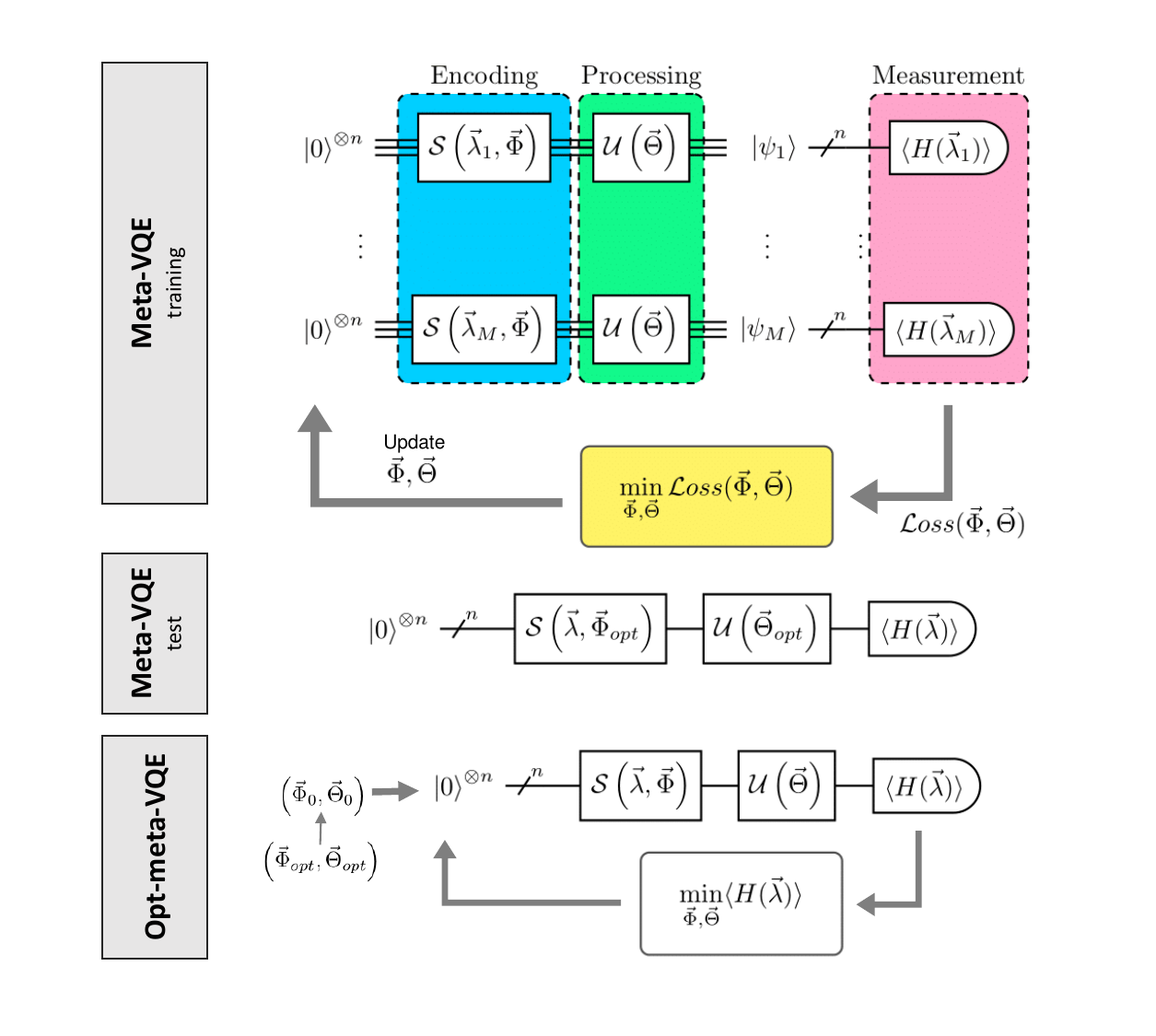}
\label{Fig:metaVQE}
\caption{The top diagram represents the training part of the algorithm, where the Hamiltonian parameters  $\vec{\lambda}$ are encoded together with the variational parameters $\vec{\Phi}$ and $\vec{\Theta}$. By computing the expected value of the Hamiltonian for multiple values of $\vec{\lambda}$, we design a cost function $\mathcal{L}oss$ to be minimized. Once the algorithm converges to a $\Phi_{opt}$ and $\Theta_{opt}$, we can use these values to obtain any $\langle H(\vec{\lambda})\rangle$ (meta-VQE test) or as initial values of a standard VQE algorithm (opt-meta-VQE).}
\end{figure*}

There are only a few techniques that try to find  quantum circuits for general Hamiltonians. Some proposals focus on finding parameterized circuits that can more efficiently explore a region of the Hilbert space \cite{sim_expressibility_2019, nakaji_expressibility_2020}. In general, all these circuit ansatz consist of smaller circuit layers that share a common structure, in such a way that these layers are concatenated until the desired precision is achieved. Although this structure may seem efficient, the exponentially large Hilbert space hampers the ability to explore it in a reasonable number of steps. Only if there exists some initial-state guess can the algorithm converge to the correct solution. The lack of this initial estimation sometimes imposes random initialization, which leads to the barren-plateau problem: both the gradient and the variance of the circuit parameters tend to zero exponentially, causing the algorithm
to get stuck in some local minimum \cite{mcclean_barren_2018}.

In the best-case scenario, with an initial-state preparation and a physically inspired circuit, a VQE run will give only a specific ground-state energy. This is in many cases insufficient when the Hamiltonian depends on some parameters (nuclear coordinates, external field strengths, model-specific parameters) and the goal is to find the configuration that leads to specific properties of the system. Some exam- ples are the lowest ground-state energy with respect to the parameters of the Hamiltonian (in molecular cases this is often referred to as geometry optimization), energy gaps between ground and excited states, the convergence behaviour of specific parameterized circuits with respect to the parameters of model systems, and, in general, the evolution of some observable with respect to an external parameter such as the strength of an electromagnetic field. Then, one has to run many instances of the VQE to scan over these parameters, inevitably increasing the computational cost. Previous work has explored the possibility of using a VQA to predict the ground state of a Hamiltonian. In particular, Ref. \cite{genVQE} proposed to use adiabatic state preparation to design a circuit ansatz capable of doing that task. However, that proves costly when more than one Trotter step is considered. We aim to generalize this idea further by using short-depth quantum circuits, and analyze different Hamiltonians and encoding strategies.

In this work, we address the general problems stated above at once by proposing the \textit{meta-VQE} algorithm. A meta-VQE encodes the Hamiltonian parameters into the first layers of the quantum circuit, dividing the circuit into two parts: encoding and processing. Next, the meta-VQE is trained with a small set of Hamiltonian parameters by constructing a cost func- tion that is a combination of all expected values. Finally, the meta-VQE has “learned” the Hamiltonian, and we can simply introduce other values of the Hamiltonian parameters into the circuit and obtain a good estimation of the ground-state energy. If this estimation is not precise enough, we can use the resulting circuit of the meta-VQE as a starting point for a standard VQE, providing a good initial guess and avoiding the random-initialization problem, i.e., barren plateaus \cite{mcclean_barren_2018}. 

This algorithm is inspired by quantum (QML) and other algorithms that use metatechniques \cite{Cao2017,Romero2017,schuld_quantum_2018, verdon_learning_2019,schuld_circuit-centric_2020,perez-salinas_data_2020,abbas_quantum_2020, cVQE}. As these algorithms propose, we design a parameterized quantum circuit to be trained with a set of values from a given model, in our case, a physical Hamiltonian. We treat the encoding part of the meta-VQE as a quan- tum neural network that learns the encoding of the Hamiltonian. The processing part guides the encoded state towards the ground state. We observe two advantages of this algorithm: \textit{(i)} it can be used to first explore the ground state energies of ground-state energies of the Hamiltonian parameter space with only a few training points and then use the result as an initial state for a precise VQE, and \textit{(ii)} the encoding in a VQA proves valuable and helps these algorithms to find the ground state. The meta-VQE can be interpreted as a QML application for quantum simulation suited for the NISQ era.
 
We present the results of this work in the following section. First, we introduce the meta-VQE algorithm from a general perspective. Next, we run a meta-VQE with a  spin-Hamiltonian example, the one-dimensional (1D) XXZ model, and then with a molecular-Hamiltonian example, the $H_{4}$ complex, consisting of two H$_2$ molecules with a fixed bond distance of 1.23{\AA} in a rectangular arrangement with a varying intermolecular distance $d$ (the same system as used in Ref. \cite{lee2018generalized}). We compare the performance of the meta-VQE with that of a standard VQE with random initialization and a standard VQE initialized with the trained parameters obtained from the meta-VQE (\textit{opt-meta-VQE}). Finally, we apply this algorithm to a state-of-the-art application: the simulation of transmon qubits in a quantum computer \cite{QCAD}. 
We discuss the results and propose further improvements of this algorithm in the Discussion section. Results where obtained via implementation of the meta-VQE protocol using \textsc{tequila}~\cite{tequila} and choosing \textsc{qulacs}~\cite{qulacs} as a quantum backend. \textsc{tequila} represents quantum objectives as generalized functions of abstract expectation values allowing arbitrary transformations and arithmetic operations on those data types in a blackboard fashion. This allows straightforward construction of meta-VQEs.

\section{Results}

\subsection{Meta-VQE algorithm}

The structure of the meta-VQE is shown diagrammatically in Fig. \ref{Fig:metaVQE}. Given an $n$ qubits parametrized Hamiltonian of the form $H = H(\vec{\lambda})$, where $\vec{\lambda} = \{\lambda_{1},\lambda_{2},\cdots,\lambda_{q}\}$ are the $q$ different parameters, we select $M$ sets of $\vec{\lambda}$ that we will use as a training set. The circuit is initialized in the $|0\rangle^{\otimes n}$ state or in the Hartree-Fock state for molecules (see next section). The first part of the circuit is the encoding unitary $\mathcal{S}$ containing parametrized gates which arguments include the Hamiltonian parameters of one of the training points $\vec{\lambda}_{i}$ and variational parameters $\vec{\Phi}$. The second unitary of the circuit $\mathcal{U}$ also contains parametrized gates that depend on $\vec{\Theta}$ variables, but not the Hamiltonian parameters. The final state of the circuit can be written as
\begin{equation}
|\psi_{i}\rangle \equiv |\psi(\vec{\lambda}_{i},\vec{\Phi},\vec{\Theta})\rangle = \mathcal{U}(\vec{\Theta})\mathcal{S}(\vec{\lambda}_{i},\vec{\Phi})|0\rangle^{\otimes n}.
\end{equation}
The meta-VQE is optimized over a set of training points $\lambda_i$ by minimizing a cost function that depends on all expected values of the Hamiltonians $H(\lambda_i)$. Here we employ a simple cost function that is the sum of all expected values
\begin{equation}
\mathcal{L}oss(\vec{\Phi},\vec{\Theta}) = \sum_{i=1}^{M}\langle\psi_{i}\lvert H(\lambda_{i})|\psi_{i}\rangle,\label{eq:cost_function_simple}
\end{equation}
but other more sophisticated or problem-dependent ones can be developed and conveniently implemented within \textsc{tequila}. Finally, we minimize this loss function with a gradient-based method obtaining the optimal values for the variational parameters, $\vec{\Phi}_{opt}$ and $\vec{\Theta}_{opt}$.

Once the meta-VQE circuit is trained, we can proceed to test its performance. We compute the expected value of the Hamiltonian with other values of $\vec{\lambda}$ by running the meta-VQE circuit with the trained parameters $\vec{\Phi}_{opt}$ and $\vec{\Theta}_{opt}$ and the corresponding $\vec{\lambda}$ values for the test point. The results show that a meta-VQE is capable of learning the profile of the ground state energy as a function of the parameters $\vec{\lambda}$, but its accuracy depends strongly on the ansatz for the encoding and processing circuit. For that reason, we also propose to use the result of the meta-VQE as an initial-state guess for a standard VQE, giving what we call \emph{opt-meta-VQE}. In the following subsections, we present particular examples of meta-VQE circuits and test them with different kinda of physical Hamiltonians.

\subsection{Spin Hamiltonian: 1D XXZ model}

\begin{figure*}[t!]
\centering
\includegraphics[width=0.8\textwidth]{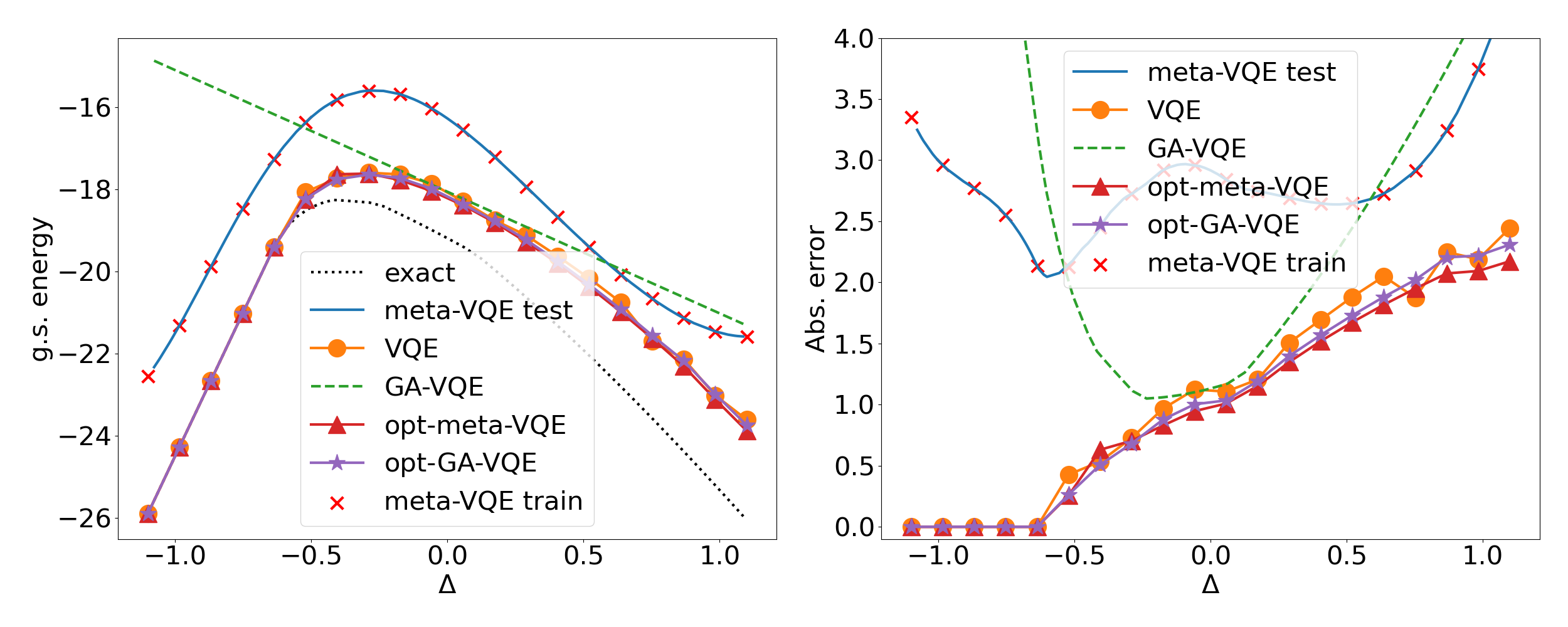}
\caption{Results for the ground state energy (arbitrary units) in relation to the $\Delta$ parameters for the $n=14$ XXZ spin chain with transverse field $\lambda=0.75$. The total number of layers considered is 4, divided into encoding and processing for meta-VQE and opt-meta-VQE. The meta-VQE learns the energy profile whereas the standard VQE achieves better precision. However, using the results of meta-VQE as a starting point of a VQE (opt-meta-VQE and opt-GA-VQE) improves notably the performance and avoids local minima, as shown by the absolute error plot with respect to the exact solution.}
\label{Fig:XXZn14}
\end{figure*}

Let us start with the 1D antiferromagnetic XXZ spin Hamiltonian with an external magnetic field
\begin{equation}
\mathcal{H}=\sum_{i=1}^{n}\sigma_{i}^{x}\sigma_{i+1}^{x}+\sigma_{i}^{y}\sigma_{i+1}^{y}+\Delta\sigma_{i}^{z}\sigma_{i+1}^{z} + \lambda \sum_{i=1}^{n}\sigma_{i}^{z},
\end{equation}
where $\Delta$ is the anisotropy parameter and $\lambda$ the transverse field strength. We also consider periodic boundary conditions. For $\lambda = 0$, this model contains two quantum phase transitions, at $\Delta = \pm 1$. Its ground state is a product state for $\Delta < -1$ and is highly entangled in the critical region, i.e. for $-1 <\Delta \leq 1$. The introduction of the external field $\lambda$ moves the $\Delta = -1$ phase transition to higher values of $\Delta$, increasing the region with a product state as a ground state \cite{langari_phase_1998}.

We use this model as a test for various reasons. First, this is a non-trivial model, with a highly entangled ground state, a property that motivates the use of a quantum computer. Second, there is no known quantum circuit capable of computing the ground state exactly, in contrast with other integrable models \cite{kraus_compressed_2011,hebenstreit_compressed_2017, schmoll_kitaev_2017,verstraete_quantum_2009,cervera-lierta_exact_2018,montanaro_compressed_2020}, although there has been recent work on the preparation of the Bethe-ansatz eigenstates using projective algorithms \cite{van2021preparing}. Third, the energy profile as a function of $\Delta$ for a non-zero value of $\lambda$ has a non-trivial shape with a peak (see Fig.~\ref{Fig:XXZn14}). Finally, some condensed matter models such as Haldane-Shastry \cite{HS-XXZ} and some electronic Hamiltonians, are related to this model.

As stated above, no known quantum circuit represents the ground state of this model for any $n$ and $\Delta$. This gives us a good training ground to test the most general approach to a meta-VQE. We follow the same strategy as in other VQAs when applied to a general problem: we construct the quantum circuit with minimal blocks, called layers, that are concatenated and progressively introduce more variational parameters and generate entanglement. One may expect that, the greater the number of layers that are considered, the closer to the ground state the algorithm will end up. The encoding part of the meta-VQE follows a similar strategy as data re-uploading for variational quantum classification \cite{perez-salinas_data_2020}: we encode the parameter $\Delta$ into rotational single-qubit gates using a linear function of the form $w\Delta + \phi$, where $w$ and $\phi$ are the variational parameters. Universality of this kind of encoding, when provided with enough layers with a single-qubit, has been shown \cite{perez-salinas_data_2020}, so we expect to obtain a similar behavior when it is used on multiple qubits and entanglement is added between them. Each layer, in both the encoding and the processing part, contains first-neighbor controlled-NOT (CNOT) gates. This kind of entanglement ansatz has been proven to provide circuits with high expressibility \cite{sim_expressibility_2019}.

An encoding layer $l$ of the meta-VQE circuit for this model can be written in the following form
\begin{multline}
S_{l}\equiv S(\Delta,\vec{\Phi}_{l})=  R\left(f(\Delta,\vec{\varphi}_{1l}))\otimes \cdots \otimes R(f(\Delta,\vec{\varphi}_{nl})\right)\\
\otimes CNOT\otimes \cdots \otimes CNOT,\label{eq:encoding_circuit}
\end{multline}
where $\vec{\Phi}_{l} = (\vec{\varphi}_{1l},\cdots, \vec{\varphi}_{nl})$ and 
\begin{equation}
R(f(\Delta,\vec{\varphi}_{il})) = R_{z}(w_{il}^{(1)}\Delta + \phi_{il}^{(1)}) R_{y}(w_{il}^{(2)}\Delta + \phi_{il}^{(2)})
\label{eq:encoding}
\end{equation}
are single-qubit rotational gates with $\vec{\varphi}_{il} = (\vec{w}_{il},\vec{\phi}_{il})$, $\vec{w}_{il} = (w_{il}^{(1)},w_{il}^{(2)})$ and $\vec{\phi}_{il} = (\phi_{il}^{(1)},\phi_{il}^{(2)})$. Then, each encoding layer contains $4n$ variational parameters. Notice that this linear encoding is similar to the classical neural-network encoding, where $\vec{w}_{il}$ and $\vec{\phi}_{il}$ play the role of the weights and biases respectively, and the rotational gate plays the role of the non-linear activation function. The encoding gate is then constructed with $L_{1}$ layers:
\begin{equation}
\mathcal{S}(\Delta,\vec{\Phi}) = S_{1}\otimes \cdots \otimes S_{L_{1}},
\end{equation}
where $\vec{\Phi}=(\vec{\Phi}_{1},\cdots,\vec{\Phi}_{L_{1}})$.

Each processing layer is constructed in the same way as in Eq.~\eqref{eq:encoding_circuit}, now with rotations that do not depend on the meta parameters. Thus, each processing layer can be written as
\begin{multline}
U_{l}\equiv U(\vec{\Theta}_{l})= R(\vec{\theta}_{1l})\otimes \cdots \otimes R(\vec{\theta}_{nl})\\
\otimes CNOT\otimes \cdots \otimes CNOT,
\label{eq:U_proc}
\end{multline}
where
\begin{equation}
R(\vec{\theta}_{il}) = R_{z}(\theta_{il}^{(1)}) R_{y}(\theta_{il}^{(2)}).
\end{equation}
Each layer of processing unitary contains $2n$ variational parameters, the components of $\vec{\theta}_{il}$. Considering $L_{2}$ processing layers, the total unitary becomes
\begin{equation}
\mathcal{U}(\vec{\Theta}) = U_{1}\otimes \cdots \otimes U_{L_{2}},
\end{equation}
with $\vec{\Theta} = (\vec{\Theta}_{1},\cdots, \vec{\Theta}_{L_{2}})$.

\begin{figure*}[t!]
\centering
\includegraphics[width=0.8\linewidth]{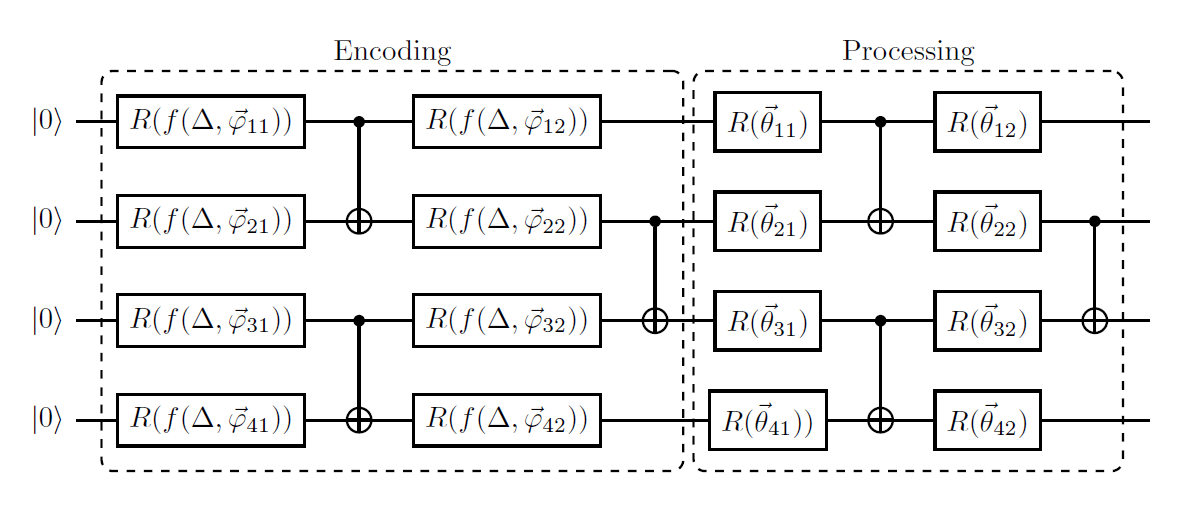}
\caption{Circuit ansatz corresponding to four qubits and two encoding and processing layers. This ansatz is used for the XXZ meta-VQE. Each $R(\vec{x})$ gate corresponds to $R_{z}(x^{(1)})R_{y}(x^{(2)})$. The function used for the encoding layer is $f(\Delta,\vec{\varphi})= w\Delta + \phi$.}
\label{fig:circuit_ansatz}
\end{figure*}

Figure \ref{fig:circuit_ansatz} shows the circuit ansatz described above for $n=4$ qubits and two encoding and processing layers.

We run the meta-VQE for the XXZ model with a transverse field $\lambda=0.75$. The training set of $\Delta$ points is composed of 20 equispaced points between $\Delta=-1.1$ and $\Delta=1.1$. The performance of the 
meta-VQE is then tested through evaluation on 100 equispaced testing points between the same $\Delta$ values (and using the parameter values found after the training). 

To check whether the encoding strategy of the meta-VQE entails an advantage, we run a meta-VQE with no encoding layers, i.e. all layers have the form of $U_{l}$ in Eq. \eqref{eq:U_proc} (a globally-averaged VQE, \emph{GA-VQE}). The number of optimization parameters is then lower than the original meta-VQE, so we keep the same total number of layers to maintain the circuit depth. 

To check the possible advantage of the meta-VQE learning strategy, we compare the results with those from a standard VQE. Again, we keep the total number of layers of $L_{1}+L_{2}$ and all layers are of the form of $U_{l}$ in Eq. \eqref{eq:U_proc}. We do not consider a single-point VQE with an encoding part because the encoding function $w\Delta + \phi$ can be expressed with a single angle, $\theta$, reducing the circuit to a standard VQE. Encoding only has an effect only in the simultaneous optimization of different points. In total, the meta-VQE and opt-meta-VQE need $n(4L_{1} + 2L_{2})$ variational parameters and GA-VQE and the standard VQE need $n(2L_{1} + 2L_{2})$.

Finally, we check whether we can use the results of a meta-VQE and GA-VQE, that is, with and without the encoding strategy respectively, as starting points for a standard VQE (\emph{opt-meta-VQE} and \emph{opt-GA-VQE}). We run the simulation of a VQE again using as initial parameters the results of the meta-VQE and of the meta-VQE without encoding (GA-VQE). In both cases, the circuit depth is the same but the optimization parameters are different. The latter initializes the quantum circuit to the same state, regardless of the Hamiltonian parameter while the former provides a parameter-dependent initialization. In contrast to the standard VQE with encoding, in this case the encoding helps us to start at a specific point: if the initialization is random, the encoding can be reduced to a single rotational parameter, but if it is not, it can be used to help the algorithm to start at a particular point. 

Figure \ref{Fig:XXZn14} shows the results of these four simulations for an XXZ spin chain of 14 qubits, with two encoding and two processing layers. In all cases, the circuit is not good enough to find the ground state energy for all values of $\Delta$. The reason is simple: the circuit ansatz does not generate enough entanglement and the rotational gates are not expressible enough. This is something that we should expect since we are considering just a few layers and in turn, their design is not physically inspired. As we mentioned before, we chose this model as an example of a worst-case scenario for a VQE-type simulation, where we have no clue regarding the circuit ansatz and the initial-state preparation. The results are much better for $\Delta \lesssim 0.6$ because the ground state there corresponds to all spins being aligned with the external magnetic field (a ferromagnetic phase), i.e. in the computational basis, the state $|11\cdots 1\rangle$. This basis element is easy to find for the algorithms due to the full basis superposition induced by the $R_{y}$ gates.

It seems that a meta-VQE can find the general energy shape but not provide an accurate value, in contrast to the standard VQE. GA-VQE can reproduce only a linear profile, not learning the behavior of the energy function. However, opt-meta-VQE proves valuable, achieving better results than standard VQE with random initialization. This result is better shown in Fig. \ref{Fig:XXZn14} right, where we plot the absolute error of all algorithms with respect to the exact solution. Whereas the standard VQE gets stuck in local minima for some values of $\Delta$, opt-meta-VQE is able to deal with those false minima and achieve better precision. 

\begin{figure}[t!]
\centering
\includegraphics[width=\columnwidth]{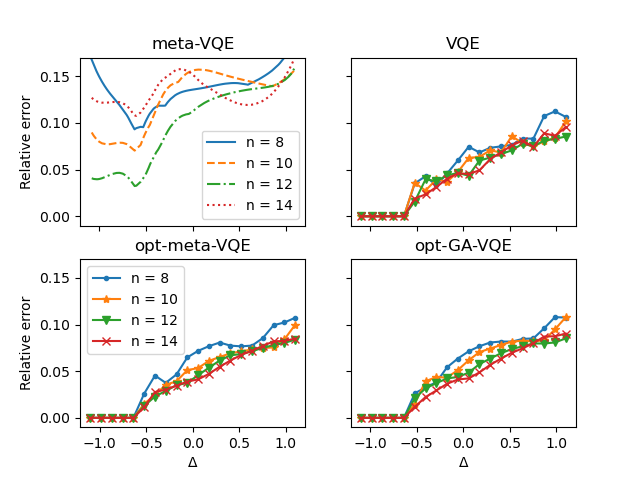}
\caption{Relative errors with respect to the exact ground state of the 1D XXZ model with $\lambda=0.75$ for the four algorithms considered in this work in relation to the $\Delta$ parameter and the number of qubits $n$. The errors for the VQE, opt-meta-VQE and opt-GA-VQE decrease slightly with the number of qubits. For the meta-VQE, the error scaling with $n$ depends on the phase of the system, a trait that can be explained by the hardness of learning a critical system with a single global cost function.}
\label{Fig:error_qubits}
\end{figure}

Finally, we compare the scaling of the results for different numbers of qubits and, therefore, an increasing number of optimization parameters. Figure \ref{Fig:error_qubits} shows the relative error with respect to the exact ground state energy for the four algorithms considered for $n=8,10,12,14$. The errors decrease slightly with the number of qubits for opt-meta-VQE and opt-GA-VQE. However, the meta-VQE shows fluctuations depending on the phase of the system and the number of qubits. This behavior may be a consequence of the hardness of the optimization; we are trying to learn the ground state energy of a critical system that undergoes to a quantum phase transition with a single global cost function. The ground state will change drastically when the critical point is crossed and the encoding may be dependent on the phase. As mentioned before, we chose this model as an example of a  hard-case. 

\subsection{Molecular Hamiltonian: H$_4$}

To explore the role of the encoding function, we employ an example of an electronic structure problem in quantum chemistry. As before, the objective is to approximate ground-states of qubit Hamiltonians, generated in this example, by transforming second-quantized fermionic operators using various transforms \cite{cao_quantum_2019, mcardle2020quantumreview}. We use the $k$-UpCCGSD approach of Ref.~\cite{lee2018generalized} in a separated form, equivalent to a single Trotter step of the full approach, as a VQE ansatz. This denotes a unitary coupled-cluster ansatz with generalized electronic single and double excitations where the double excitations are limited to electron pairs within the same spatial orbital. This ansatz is repeated $k$ times, similarly to the layered construction used before. When applied to a reference state, unitary coupled-cluster approaches are number conserving, which is a necessary condition for electronic eigenstates. In this work we use the Hartree-Fock (best product state from a classical mean-field optimization) state as a reference. In contrast to the spin Hamiltonians above the circuit construction is physically motivated illustrating a different use case of the meta-VQE approach. In the next paragraphs we give a brief description on the construction of the parametrized quantum circuit following the implementations presented in Refs.~\cite{kottmann2021feasible, tequila}.

The overall circuit consists of a collection of primitive unitary coupled-cluster operations that describe collective excitations of $n$-electrons from spin-orbitals $\mathbf{p}=(p_0,p_1,\dots,p_n)$ into spin-orbitals $\mathbf{q}=(q_0,q_1,\dots,q_n)$
\begin{equation}
U\left(\theta_{\mathbf{p}\mathbf{q}}\right) = e^{-i\frac{\theta_{\mathbf{p}\mathbf{q}}}{2} G_{\mathbf{p}\mathbf{q}}}.
\end{equation}
The generators $G_{\mathbf{p}\mathbf{q}}$ are constructed from pairs of fermionic creation and annihilation operators and mapped to linear combinations of Pauli strings by various transformations with the most prominent being the Jordan-Wigner transformation which we use in this paper.
In the case of $k$-UpCCGSD the types of excitation are restricted to single and double excitations where the corresponding spin-orbitals $p_0,p_1,q_0,q_1$ are restricted to the same spatial orbital. If we index the spatial orbitals with $P$ and label the spin-up (spin-down) orbitals with even (odd) numbers, a pair restricted excitation generator in the Jordan-Wigner representation can be written as
\begin{multline}
    G_{(2P,2P+1),(2Q, 2Q+1)} \\ \xrightarrow{JW} i\left(\sigma^+_{2P}\sigma^-_{2Q}\sigma^+_{2P+1}\sigma^-_{2Q+1} - h.c.\right),
\end{multline}
 where $\sigma^{\pm}=\left(\sigma^{x}\pm i\sigma^{y}\right)/2$.
The single excitations behave similarly but introduce additional $\sigma^z$ terms on intermediate qubits
\begin{multline}
    G_{2P,2Q} \xrightarrow{JW} i\left( \sigma^+_{2P} \left(\prod_{k=2P+1}^{2Q-1}\sigma^z_k\right) \sigma^-_{2Q} - h.c. \right), \end{multline}
and analogously for $G_{2P+1, 2Q+1}$.
A single layer of the factorized $k$-UpCCGSD circuit is then constructed from all possible restricted double excitations and the corresponding single excitations excluding spin flips
\begin{multline}
    U_\text{pCCGSD}\left(\boldsymbol{\theta}\right) = \\ \prod_{P<Q}\left(\phantom{\prod_{P,Q}}  e^{-i\frac{\theta_{(2P,2Q),(2P+1,2Q+1)}}{2} G_{(2P,2Q),(2P+1,2Q+1)}} \right.\\ \left. e^{-i\frac{\theta_{2P,2Q}}{2} G_{2P,2Q}}  \phantom{\prod_{P,Q}} e^{-i\frac{\theta_{2P+1,2Q+1}}{2} G_{2P+1,2Q+1}} \right),
\end{multline}
with $\boldsymbol{\theta}=\left(\theta_{0213},\theta_{02}, \theta_{13},\cdots,\theta_{2P,2Q,2P+1,2Q+1},\cdots\right)$, following the implementation in Ref.~\cite{tequila}.
The full $k$-UpCCGSD circuit consists of $k$ layers with individual parameter sets acting on an initial state commonly chosen as the Hartree-Fock state 
\begin{equation}
    U\left(\boldsymbol{\theta}\right) = \prod_k U_{\text{pCCGSD}}\left(\boldsymbol{\theta}^{(k)}\right) U_\text{HF}.
    \label{eq:ansatz-k-upccgsd}
    \end{equation}
The unitary $U_\text{HF}$ prepares the Hartree-Fock state which is the fermionic state with the first $N_\text{electrons}$ molecular spin-orbitals occupied. In the Jordan-Wigner encoding this translates to a simple computational basis state
\begin{equation}
    U_\text{HF} \xrightarrow{JW} \prod_{l=0}^{{N_\text{electrons}}} \sigma^x_{l}.
\end{equation}

In the standard meta-VQE approach, each UpCCGSD angle is encoded in the same way as in the previous sections using a meta parameter $d$ and a linear function,
\begin{align}
    \theta_{\mathbf{p}\mathbf{q}} = w_{\mathbf{p}\mathbf{q}} \ d + \phi_{\mathbf{p}\mathbf{q}},
    \label{eq:upccgsd-meta-encoding}
\end{align}
with independent variational parameters $w_{\mathbf{pq}}$ and $\phi_{\mathbf{pq}}$.
Additionally we introduce a non-linear encoding (\emph{nl-meta-VQE}) using a single Gaussian to encode each UpCCGSD angle as
\begin{equation}
    \theta_{\mathbf{p}\mathbf{q}} = \alpha_{\mathbf{p}\mathbf{q}} e^{\beta_{\mathbf{p}\mathbf{q}}\left(\gamma_{\mathbf{p}\mathbf{q}} - d \right)} + \delta_{\mathbf{p}\mathbf{q}},
    \label{eq:upccgsd-nl-meta-encoding}
\end{equation}
with individual variational parameters $\alpha_{\mathbf{pq}}$, $\beta_{\mathbf{pq}}$, $\gamma_{\mathbf{pq}}$ and $\delta_{\mathbf{pq}}$.
In comparison with the previous sections and Fig.~\ref{Fig:metaVQE} this approach contains a static (non-parametrized) initial part $U_\text{HF}$ and an encoded part $U_\text{pCCGSD}$. Note that, although it could be envisioned in future approaches, there is no processing part in the examples explored here.

\begin{figure}
    \includegraphics[width=0.45\textwidth]{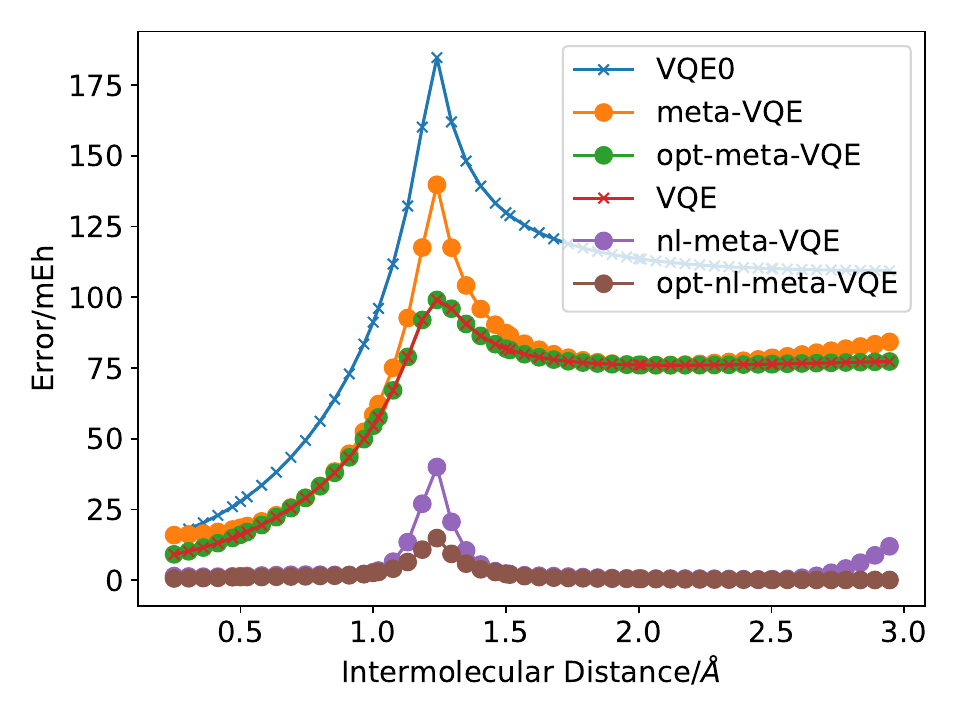}
    \caption{Relative errors for the rectangular H$_4$ molecule in eight spin-orbitals (STO-3G basis set) in relation to the intermolecular bond distance $d$. The 2-UpCCGSD model was used for all VQEs. The VQE lines denote the standard optimized 2-UpCCGSD model starting from the Hartree-Fock configuration (VQE0). Linear encoding (meta-VQE) replaces the UpCCGSD angles with an encoding of the form $\theta = \alpha + d\beta $, while the non-linear encoding (nl-meta-VQE) uses a floating Gaussian as $\theta = \alpha e^{\beta\left(\gamma - d \right)} + \delta $. The opt-meta-VQE line denotes the canonical VQE initialized with the angles from the meta-VQE. All errors are with respect to the exact diagonalization of the Hamiltonian within this basis set. Training points are chosen from 0.5{\AA} to 2.5{\AA} with 0.5{\AA} distance between them. 
    }\label{fig:h4_example}
\end{figure}

In Fig.~\ref{fig:h4_example}, we show the results of the meta-VQE with $2$-UpCCGSD applied to the dimer of two hydrogen molecules in a rectangular arrangement (see for example Ref.~\cite{lee2018generalized}) using the distance between the two molecules as a meta parameter. At 1.23{\AA}, this results in a quadratic structure and degeneracies within the orbitals due to a change in the point-group symmetry making this point challenging for standard single-reference approaches in classical quantum chemistry. 

A single layer of the UpCCGSD ansatz is not able to achieve the accuracies below the milliHartree threshold, in both absolute and relative energies, that are required for accurate chemical predictions. Using more than one layer of UpCCGSD gives enough additional freedom to the circuit, but the usual optimization strategy that initializes all angles to zero (i.e. starting from the reference state) fails to converge here and leads to no improvement from the additional layers. The meta-VQE is able to capture the basic form of the VQE results at the individual points and reaches similar accuracy as the canonical VQE at the individual points if the system is not close to the critical point.

In the original work~\cite{lee2018generalized}, repeated random initialization was used to obtain convergence to the best possible result. Here, we initialize all angles to zero for the meta-VQE as well as for the regular VQE using the BFGS optimizer of \textsc{scipy} with  settings that worked well for previously investigated molecular systems ~\cite{kottmann2020reducing}. 

We extend the meta-VQE to a non-linear encoding as given in Eq.~\eqref{eq:upccgsd-nl-meta-encoding}, abbreviated as \emph{nl-meta-VQE}, using a single Gaussian to encode each UpCCGSD angle and the intermolecular distance of the two hydrogen molecules as meta parameter $d$. The parameters are initialized such that the initial values produced are the same as for the linear meta-VQE and the canonical VQE ($\alpha,\delta = 0$, $\beta,\gamma = 1$). The non-linear encoding results in significantly improved convergence which can be further improved by individually optimizing the individual points with a regular VQE. Floating Gaussians are fairly good general function approximators and in this example using a single one is sufficient for the corresponding interval of interest. In addition, floating Gaussians are able to capture the asymptotic behavior of molecular distances over the constant offset $\delta$. It is expected that with increasing intermolecular distance $d$ the dependence of the UpCCGSD parameters should asymptotically approach a constant value since the system will resemble two isolated molecules. A representation with floating Gaussians automatically incorporates this property. In the same way this holds true for bond distances -- see our online material for a detailed example based on a single hydrogen molecule. More general non-linear encodings are currently under investigation.

In Fig.~\ref{fig:h4_example} a standard VQE takes on average 3345 evaluations of the objective function, while opt-nl-meta-VQE takes 204. The training of nl-meta-VQE takes 12760 evaluations of the objective, including all five training points. The cost of obtaining a given number of test points in terms of individual expectation value estimations can then be estimated as $12760\cdot n_\text{training-points} + 204\cdot n_\text{test-points}$ for the opt-nl-meta-VQE, and $3345\cdot n_\text{test-points}$ for the standard VQE. Since we use five training points, opt-nl-meta-VQE becomes cheaper when more than 20 test points are required.
Note that those numbers should not be interpreted as rigorous benchmarks. They reflect the state of our implementation in \textsc{tequila} at the time of writing of manuscript. We expect, for example, further improvement by incorporating analytical gradients for unitary coupled-cluster methods~\cite{kottmann2021feasible} in combination with an improved automatic differentiation scheme, that avoids multiple evaluations of the same objective.

\subsection{Transmon simulation}

As a final example, we apply the meta-VQE to a state-of-the-art application of quantum computation: the quantum computer-aided design (QCAD) of superconducting digital quantum computers \cite{QCAD}. We use this example to show how the meta-VQE can be applied straightforwardly to recent work. 

We consider the truncated four-qubit Gray-encoded \cite{d-level} Hamiltonian for a flux-tunable transmon presented in Ref. \cite{QCAD}. Following the original reference, as a circuit ansatz for both the encoding and the processing layer of the meta-VQE, we consider single-qubit rotations $R_{x}$ and $R_{z}$ applied on each qubit and parameterized $XX$ gates applied to all six pairs of qubits. As a further simplification, the parameters of the $XX$ gates are common to all layers, leading to the same entangling gate after each layer of single-qubit rotations. Finally, we add a layer of $R_{x}$ and $R_{z}$ gates at the end of the processing layer. We encode linearly the Hamiltonian parameter $f$ (flux) in only the rotational gates of the encoding layers, applying the same encoding strategy as for the XXZ Hamiltonian. 

In this example, we compare the meta-VQE and opt-meta-VQE with  the standard VQE and improved versions of the VQE. 
In particular, we define \textit{VQE-smart} as a standard VQE that uses as initial points the optimized points of the previous Hamiltonian parameter optimization. This approach assumes that the ground state does not change when moving drastically from close points in the Hamiltonian parameter space. We also consider the effect of the encoding layers in this kind of VQE. We define \textit{VQE-enc} as a VQE that has the same parameterized quantum circuit as in the meta-VQE and in opt-meta-VQE but uses as the starting point of each minimization the optimized variables of the previous point, as in VQE-smart. The main difference between VQE-enc and opt-meta-VQE is the choice of the initialization parameters: the result of the previous optimization at each step and the result of the meta-VQE, respectively.
This methodology can be further improved by running, for instance, layerwise optimization strategies \cite{lw_VQE}. 

\begin{figure*}
\centering
\includegraphics[width=0.8\textwidth]{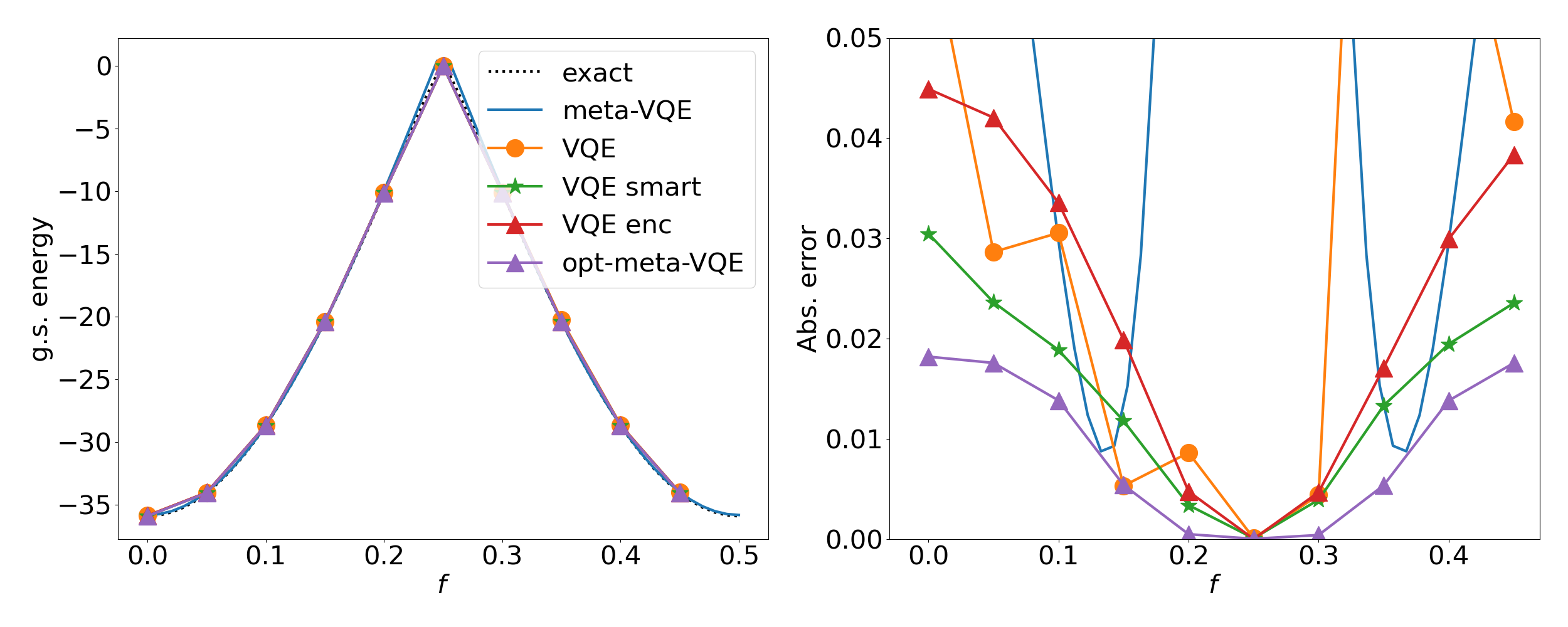}
\caption{Left: ground state energy as a function of the normalized external magnetic flux $f$ for the QCAD~\cite{QCAD} simulation of a single transmon. Right: absolute error with respect to the exact diagonalization of the Hamiltonian. We compare the meta-VQE and opt-meta-VQE with VQE and improved versions of it, where we use the result of the previous optimization point to initialize the parameters of the next one (\textit{VQE-smart} and the VQE with encoding layers (\textit{VQE-enc}). 
Both the VQE-smart and the VQE-enc show more consistent results in comparison with VQE with the random initialization, where the algorithm gets trapped in local minima for some points. The introduction of encoding layers in VQE-enc seems not to be beneficial in comparison with the standard VQE circuit (VQE-smart). This could be a consequence of the increase in the number of parameters, where the optimization can face the barren-plateau problem \cite{mcclean_barren_2018}. On the other hand, opt-meta-VQE shows the best performance, indicating that the use of the meta-VQE result as the initialization point better captures the parameter-space region of the ground state.}
\label{fig:transmon_results}
\end{figure*}

The results for one encoding and one processing layer are shown in Fig.~\ref{fig:transmon_results}. The meta-VQE learns the ground-state energy with high accuracy using only 10 training points. The performance of opt-meta-VQE is better than that of the VQE, VQE-smart and VQE-enc. As one might expect, the VQE with random initialization gets stuck in local minima for some of the test points. On the other hand, VQE-smart and VQE-enc show more consistent results. The encoding strategy in VQE-enc does not prove beneficial for increasing the precision. The increase in the number of parameters to be optimized in comparison with VQE-smart can produce the emergence of local minima or barren-plateaus \cite{mcclean_barren_2018}, i.e. the gradients are close to zero making it difficult to explore the parameter space. However, opt-meta-VQE shows the most precise results. This fact implies that a good initialization is crucial to find the best minimum. In this case, using the initial parameters provided by the meta-VQE outperforms the initialization strategy of using the previously optimized parameters as starting point, maybe because the assumption that the ground state does not change significantly from one point to another is not accurate enough.

Besides the better performance of opt-meta-VQE, notice that for the VQE and the VQE with encoding, to obtain any point of the g.s. energy, one needs to first optimize a close point and then use the result to optimize the target point. With opt-meta-VQE this is not necessary, and any point can be obtained by using the result of the meta-VQE. The number of optimization variables involved is 30 for the VQE and 38 for the rest (the meta-VQE, opt-meta-VQE, and the VQE with encoding).

\section{Discussion}

We present the meta-VQE, an algorithm to compute the ground state energy of a parametrized Hamiltonian. In contrast to standard VQE, meta-VQE learns the Hamiltonian ground-state structure and delivers an initial circuit parametrization that can be used to compute any ground state energy. To do so, the meta-VQE circuit is divided into two parts, the encoding and processing unitaries. First, this circuit is trained with a set of Hamiltonian parameters, which are encoded in the gates of the encoding unitary. By designing a cost function with the expected values of all these Hamiltonian training points, the algorithm extracts the optimal values of the variational parameters. Then, we can compute the energy for other Hamiltonian values by just running the meta-VQE circuit with the parameters obtained in the minimization. In addition, we can also use the result of a meta-VQE training process as a starting point for a standard VQE algorithm, referred to as opt-meta-VQE, instead of using random initialization. The meta-VQE captures global correlation with a few training points alleviating the need for refined optimization of the individual points in a later step. With this approach, we do not intend primarily to find stationary points on the potential-energy surface but instead to capture the dependence of the VQE model parameters with meta-parameters that appear in the Hamiltonian in specific subspaces.

We demonstrate the efficacy of the algorithm numerically by simulating a spin chain Hamiltonian, the XXZ model, the electronic Hamiltonian of H$_{4}$, and a single-transmon Hamiltonian, representing different scenarios for future applications. In the first example, we run a meta-VQE for an $n=14$ XXZ spin chain, with two encoding and processing layers and 10 training points. We compare the results with those of a standard VQE, opt-GA-VQE, and opt-meta-VQE. The results show that the meta-VQE learns the ground state energy shape but its accuracy is lower than that of a standard VQE. However, in many cases, opt-meta-VQE is more precise than the standard VQE. For the molecular case, the H$_{4}$ system illustrates an application where physically inspired circuit construction can be applied. The meta-VQE is again able to capture the basic features of the energy profile. We use this particular example to illustrate a possible extension of the meta-VQE to non-linear encodings where a simple first approach could globally improve the convergence beyond that of the standard VQE, reaching the full potential of the underlying circuit. In the third example, the meta-VQE learns the ground-state energy with high accuracy, and the performance of opt-meta-VQE is better than the standard VQE and of the VQE with encoding.

We use the XXZ spin chain as a test because of its generality. There is no known circuit ansatz to compute the ground state of this model, so the meta-VQE structure is a general one. This example represents a worst-case scenario. In the molecular case, we illustrate a scenario where circuit construction based on physical principles is possible. Finally, the single-transmon example represents a state-of-the art application and shows how the meta-VQE can be easily adapted to increase the accuracy of both well-known and state-of-the-art algorithms.

The fact that opt-meta-VQE is more precise and is able to avoid some local minima, can be interpreted as a new way to try to avoid the barren plateau problem \cite{mcclean_barren_2018} using global information. For the transmon qubit simulation example, opt-meta-VQE is significantly more accurate than other initialization strategies. The meta-VQE approach can be added to with other solutions to this problem by, for instance, using local cost functions \cite{cerezo_cost-function-dependent_2020,premaratne_engineering_2020}, adding correlations between the circuit parameters \cite{volkoff_large_2020}, using a different initialization strategy \cite{grant_initialization_2019}, or exploiting noise \cite{skolik_layerwise_2020,wang_noise-induced_2020}. All these other approaches can be adapted both to the training of the meta-VQE and to opt-meta-VQE.

We can also adapt the meta-VQE to other improved VQE models \cite{barkoutsos_quantum_2018,cowtan_generic_2020, fujii_deep_2020,huggins_non-orthogonal_2020,chivilikhin_mog-vqe_2020}. The encoding strategy also has possibilities for improvement. Classical and quantum machine learning techniques can be used to encode the Hamiltonian parameters more efficiently. In the end, the aim of this work is to show the general strategy of the meta-VQE when used in the most general way and for the most general VQE circuit. We illustrate the potential of non-linear encodings with the H$_4$ example, show how better accuracy can be obtained with the transmon simulation example, and anticipate further more generalized improvements in the future. One example is the usage of multiple meta-parameters, such as multiple bond lengths and angles, in the molecular case, or the transverse field $\lambda$ in the XXZ example. In such cases, we expect the non-linear encoding of Eq.~\eqref{eq:upccgsd-nl-meta-encoding} in the form $\theta_{\mathbf{p}\mathbf{q}} = f(d_0,d_1,\dots)$ to have significant advantages compared with the linear encoding. Following the data re-uploading strategy of Ref.~\cite{perez-salinas_data_2020}, we can also encode each meta-parameter in one single-qubit rotational gate (e.g. $\Delta$ in the $R_{z}$ gate and $\lambda$ in the $R_{y}$ in Eq.\eqref{eq:encoding}). Apart from multi-dimensional floating Gaussians, we can also envision more sophisticated approaches, such as neural networks, approximating and training the non-linear function $f$. All simulations shown in this work are exact, i.e. they do not contain any noise model. We expect the meta-VQE to suffer similar effects of noise to other VQAa, effects that can be partially compensated by the use of error-mitigation techniques \cite{suguru2018practical}.

Finally, we would like to emphasize the gain in efficiency that a meta-VQE introduces. When dealing with parametrized Hamiltonians, the only way to find the configuration with the lowest ground state energy is to scan over the parameters and run a VQE for each point, i.e. a minimization procedure for each of these points. Gradient-based minimization, as often performed for molecules, can be applied if analytical derivatives of the Hamiltonian are available, but those depend on good initial guesses of the Hamiltonian parameters. Furthermore such approaches currently are not able to exploit global information from previous points. For multi-parameter Hamiltonian studies, this reduction in the number of points to be scanned to find the interesting energy regions is even more relevant. With a meta-VQE, one can scan over the same set of parameters with a single minimization, obtain the energy profile and identify the areas of interest, e.g. those with minimal (equilibrium structures) or maximal energy (transition state search). In combination with more advanced algorithms, asuch as excited-state optimization, a meta-VQE could be used to represent the ground state in a first coarse search for points of interest, such as points with a low energy gap. This is, for example, the case for the transmon system of Ref.~\cite{QCAD} that we illustrate here as one potential use case. 
Then, if more accuracy is required, one can run an individual VQE minimization or opt-meta-VQE in these identified regions. With this approach, one can save precious computational power by avoiding minimizations in uninteresting regions of the Hamiltonian phase space. In addition, meta-VQEs can be envisioned as a valuable tool for the development of more sophisticated initialization schemes in the spirit of the molecular example investigated in this work.

\section*{Data availability}

All data used to generate the plots and extract the conclusions of this article can be found in the public repository \url{https://github.com/aspuru-guzik-group/Meta-VQE}. The main code used to obtain these data can also be found in the same repository in a tutorial form.

\section*{Acknowledgements}

We are thankful for the comments and suggestions from our Matter Lab colleagues, in particular to Thi Ha Kyaw and Sukin Sim.
This work was suppored by the U.S. Department of Energy under Award No. DE-AC02-05CH11231 (LBNL - 505736) and Award No. DE-SC0019374.
We acknowledges the generous support from Google, Inc.  in the form of a Google Focused Award. A.A.-G. also acknowledges support from the Canada Industrial Research Chairs  Program and the Canada 150 Research Chairs Program. We thank the generous support of Anders G. Fr\o{}seth the Vannevar Bush Faculty Fellowship Program.

\bibliographystyle{apsrev4-1}
\bibliography{metaVQE_bib}

\begin{thebibliography}{51}%
\makeatletter
\providecommand \@ifxundefined [1]{%
 \@ifx{#1\undefined}
}%
\providecommand \@ifnum [1]{%
 \ifnum #1\expandafter \@firstoftwo
 \else \expandafter \@secondoftwo
 \fi
}%
\providecommand \@ifx [1]{%
 \ifx #1\expandafter \@firstoftwo
 \else \expandafter \@secondoftwo
 \fi
}%
\providecommand \natexlab [1]{#1}%
\providecommand \enquote  [1]{``#1''}%
\providecommand \bibnamefont  [1]{#1}%
\providecommand \bibfnamefont [1]{#1}%
\providecommand \citenamefont [1]{#1}%
\providecommand \href@noop [0]{\@secondoftwo}%
\providecommand \href [0]{\begingroup \@sanitize@url \@href}%
\providecommand \@href[1]{\@@startlink{#1}\@@href}%
\providecommand \@@href[1]{\endgroup#1\@@endlink}%
\providecommand \@sanitize@url [0]{\catcode `\\12\catcode `\$12\catcode
  `\&12\catcode `\#12\catcode `\^12\catcode `\_12\catcode `\%12\relax}%
\providecommand \@@startlink[1]{}%
\providecommand \@@endlink[0]{}%
\providecommand \url  [0]{\begingroup\@sanitize@url \@url }%
\providecommand \@url [1]{\endgroup\@href {#1}{\urlprefix }}%
\providecommand \urlprefix  [0]{URL }%
\providecommand \Eprint [0]{\href }%
\providecommand \doibase [0]{http://dx.doi.org/}%
\providecommand \selectlanguage [0]{\@gobble}%
\providecommand \bibinfo  [0]{\@secondoftwo}%
\providecommand \bibfield  [0]{\@secondoftwo}%
\providecommand \translation [1]{[#1]}%
\providecommand \BibitemOpen [0]{}%
\providecommand \bibitemStop [0]{}%
\providecommand \bibitemNoStop [0]{.\EOS\space}%
\providecommand \EOS [0]{\spacefactor3000\relax}%
\providecommand \BibitemShut  [1]{\csname bibitem#1\endcsname}%
\let\auto@bib@innerbib\@empty
\bibitem [{\citenamefont {Preskill}(2018)}]{preskill_quantum_2018}%
  \BibitemOpen
  \bibfield  {author} {\bibinfo {author} {\bibfnamefont {J.}~\bibnamefont
  {Preskill}},\ }\href {\doibase 10.22331/q-2018-08-06-79} {\bibfield
  {journal} {\bibinfo  {journal} {Quantum}\ }\textbf {\bibinfo {volume} {2}},\
  \bibinfo {pages} {79} (\bibinfo {year} {2018})}\BibitemShut {NoStop}%
\bibitem [{\citenamefont {Bharti}\ \emph {et~al.}(2021)\citenamefont {Bharti},
  \citenamefont {Cervera-Lierta}, \citenamefont {Kyaw}, \citenamefont {Haug},
  \citenamefont {Alperin-Lea}, \citenamefont {Anand}, \citenamefont {Degroote},
  \citenamefont {Heimonen}, \citenamefont {Kottmann}, \citenamefont {Menke}
  \emph {et~al.}}]{NISQreview}%
  \BibitemOpen
  \bibfield  {author} {\bibinfo {author} {\bibfnamefont {K.}~\bibnamefont
  {Bharti}}, \bibinfo {author} {\bibfnamefont {A.}~\bibnamefont
  {Cervera-Lierta}}, \bibinfo {author} {\bibfnamefont {T.~H.}\ \bibnamefont
  {Kyaw}}, \bibinfo {author} {\bibfnamefont {T.}~\bibnamefont {Haug}}, \bibinfo
  {author} {\bibfnamefont {S.}~\bibnamefont {Alperin-Lea}}, \bibinfo {author}
  {\bibfnamefont {A.}~\bibnamefont {Anand}}, \bibinfo {author} {\bibfnamefont
  {M.}~\bibnamefont {Degroote}}, \bibinfo {author} {\bibfnamefont
  {H.}~\bibnamefont {Heimonen}}, \bibinfo {author} {\bibfnamefont {J.~S.}\
  \bibnamefont {Kottmann}}, \bibinfo {author} {\bibfnamefont {T.}~\bibnamefont
  {Menke}},  \emph {et~al.},\ }\href {https://arxiv.org/abs/2101.08448}
  {\bibfield  {journal} {\bibinfo  {journal} {arXiv:2101.08448}\ } (\bibinfo
  {year} {2021})}\BibitemShut {NoStop}%
\bibitem [{\citenamefont {Sim}\ \emph {et~al.}(2019)\citenamefont {Sim},
  \citenamefont {Johnson},\ and\ \citenamefont
  {Aspuru‐Guzik}}]{sim_expressibility_2019}%
  \BibitemOpen
  \bibfield  {author} {\bibinfo {author} {\bibfnamefont {S.}~\bibnamefont
  {Sim}}, \bibinfo {author} {\bibfnamefont {P.~D.}\ \bibnamefont {Johnson}}, \
  and\ \bibinfo {author} {\bibfnamefont {A.}~\bibnamefont {Aspuru‐Guzik}},\
  }\href {\doibase 10.1002/qute.201900070} {\bibfield  {journal} {\bibinfo
  {journal} {Adv. Quantum. Tech.}\ }\textbf {\bibinfo {volume} {2}},\ \bibinfo
  {pages} {1900070} (\bibinfo {year} {2019})}\BibitemShut {NoStop}%
\bibitem [{\citenamefont {Peruzzo}\ \emph {et~al.}(2014)\citenamefont
  {Peruzzo}, \citenamefont {McClean}, \citenamefont {Shadbolt}, \citenamefont
  {Yung}, \citenamefont {Zhou}, \citenamefont {Love}, \citenamefont
  {Aspuru-Guzik},\ and\ \citenamefont {O'Brien}}]{peruzzo_variational_2014}%
  \BibitemOpen
  \bibfield  {author} {\bibinfo {author} {\bibfnamefont {A.}~\bibnamefont
  {Peruzzo}}, \bibinfo {author} {\bibfnamefont {J.}~\bibnamefont {McClean}},
  \bibinfo {author} {\bibfnamefont {P.}~\bibnamefont {Shadbolt}}, \bibinfo
  {author} {\bibfnamefont {M.-H.}\ \bibnamefont {Yung}}, \bibinfo {author}
  {\bibfnamefont {X.-Q.}\ \bibnamefont {Zhou}}, \bibinfo {author}
  {\bibfnamefont {P.~J.}\ \bibnamefont {Love}}, \bibinfo {author}
  {\bibfnamefont {A.}~\bibnamefont {Aspuru-Guzik}}, \ and\ \bibinfo {author}
  {\bibfnamefont {J.~L.}\ \bibnamefont {O'Brien}},\ }\href {\doibase
  10.1038/ncomms5213} {\bibfield  {journal} {\bibinfo  {journal} {Nat.
  Commun.}\ }\textbf {\bibinfo {volume} {5}},\ \bibinfo {pages} {4213}
  (\bibinfo {year} {2014})}\BibitemShut {NoStop}%
\bibitem [{\citenamefont {McClean}\ \emph {et~al.}(2016)\citenamefont
  {McClean}, \citenamefont {Romero}, \citenamefont {Babbush},\ and\
  \citenamefont {Aspuru-Guzik}}]{mcclean_theory_2016}%
  \BibitemOpen
  \bibfield  {author} {\bibinfo {author} {\bibfnamefont {J.~R.}\ \bibnamefont
  {McClean}}, \bibinfo {author} {\bibfnamefont {J.}~\bibnamefont {Romero}},
  \bibinfo {author} {\bibfnamefont {R.}~\bibnamefont {Babbush}}, \ and\
  \bibinfo {author} {\bibfnamefont {A.}~\bibnamefont {Aspuru-Guzik}},\ }\href
  {\doibase 10.1088/1367-2630/18/2/023023} {\bibfield  {journal} {\bibinfo
  {journal} {New J. Phys.}\ }\textbf {\bibinfo {volume} {18}},\ \bibinfo
  {pages} {023023} (\bibinfo {year} {2016})}\BibitemShut {NoStop}%
\bibitem [{\citenamefont {McClean}\ and\ \citenamefont
  {Aspuru-Guzik}(2015)}]{mcclean_compact_2015}%
  \BibitemOpen
  \bibfield  {author} {\bibinfo {author} {\bibfnamefont {J.~R.}\ \bibnamefont
  {McClean}}\ and\ \bibinfo {author} {\bibfnamefont {A.}~\bibnamefont
  {Aspuru-Guzik}},\ }\href {\doibase 10.1039/C5RA23047K} {\bibfield  {journal}
  {\bibinfo  {journal} {RSC Adv.}\ }\textbf {\bibinfo {volume} {5}},\ \bibinfo
  {pages} {102277} (\bibinfo {year} {2015})}\BibitemShut {NoStop}%
\bibitem [{\citenamefont {Romero}\ \emph {et~al.}(2018)\citenamefont {Romero},
  \citenamefont {Babbush}, \citenamefont {McClean}, \citenamefont {Hempel},
  \citenamefont {Love},\ and\ \citenamefont {Aspuru-Guzik}}]{Romero2018}%
  \BibitemOpen
  \bibfield  {author} {\bibinfo {author} {\bibfnamefont {J.}~\bibnamefont
  {Romero}}, \bibinfo {author} {\bibfnamefont {R.}~\bibnamefont {Babbush}},
  \bibinfo {author} {\bibfnamefont {J.~R.}\ \bibnamefont {McClean}}, \bibinfo
  {author} {\bibfnamefont {C.}~\bibnamefont {Hempel}}, \bibinfo {author}
  {\bibfnamefont {P.~J.}\ \bibnamefont {Love}}, \ and\ \bibinfo {author}
  {\bibfnamefont {A.}~\bibnamefont {Aspuru-Guzik}},\ }\href {\doibase
  10.1088/2058-9565/aad3e4} {\bibfield  {journal} {\bibinfo  {journal} {Quantum
  Sci. Technol.}\ }\textbf {\bibinfo {volume} {4}},\ \bibinfo {pages} {014008}
  (\bibinfo {year} {2018})}\BibitemShut {NoStop}%
\bibitem [{\citenamefont {Cao}\ \emph {et~al.}(2019)\citenamefont {Cao},
  \citenamefont {Romero}, \citenamefont {Olson}, \citenamefont {Degroote},
  \citenamefont {Johnson}, \citenamefont {Kieferová}, \citenamefont
  {Kivlichan}, \citenamefont {Menke}, \citenamefont {Peropadre}, \citenamefont
  {Sawaya}, \citenamefont {Sim}, \citenamefont {Veis},\ and\ \citenamefont
  {Aspuru-Guzik}}]{cao_quantum_2019}%
  \BibitemOpen
  \bibfield  {author} {\bibinfo {author} {\bibfnamefont {Y.}~\bibnamefont
  {Cao}}, \bibinfo {author} {\bibfnamefont {J.}~\bibnamefont {Romero}},
  \bibinfo {author} {\bibfnamefont {J.~P.}\ \bibnamefont {Olson}}, \bibinfo
  {author} {\bibfnamefont {M.}~\bibnamefont {Degroote}}, \bibinfo {author}
  {\bibfnamefont {P.~D.}\ \bibnamefont {Johnson}}, \bibinfo {author}
  {\bibfnamefont {M.}~\bibnamefont {Kieferová}}, \bibinfo {author}
  {\bibfnamefont {I.~D.}\ \bibnamefont {Kivlichan}}, \bibinfo {author}
  {\bibfnamefont {T.}~\bibnamefont {Menke}}, \bibinfo {author} {\bibfnamefont
  {B.}~\bibnamefont {Peropadre}}, \bibinfo {author} {\bibfnamefont {N.~P.~D.}\
  \bibnamefont {Sawaya}}, \bibinfo {author} {\bibfnamefont {S.}~\bibnamefont
  {Sim}}, \bibinfo {author} {\bibfnamefont {L.}~\bibnamefont {Veis}}, \ and\
  \bibinfo {author} {\bibfnamefont {A.}~\bibnamefont {Aspuru-Guzik}},\ }\href
  {\doibase 10.1021/acs.chemrev.8b00803} {\bibfield  {journal} {\bibinfo
  {journal} {Chem. Rev.}\ }\textbf {\bibinfo {volume} {119}},\ \bibinfo {pages}
  {10856} (\bibinfo {year} {2019})}\BibitemShut {NoStop}%
\bibitem [{\citenamefont {{Yordanov}}\ \emph {et~al.}(2020)\citenamefont
  {{Yordanov}}, \citenamefont {{Arvidsson-Shukur}},\ and\ \citenamefont
  {{Barnes}}}]{yordanov_efficient_2020}%
  \BibitemOpen
  \bibfield  {author} {\bibinfo {author} {\bibfnamefont {Y.~S.}\ \bibnamefont
  {{Yordanov}}}, \bibinfo {author} {\bibfnamefont {D.~R.~M.}\ \bibnamefont
  {{Arvidsson-Shukur}}}, \ and\ \bibinfo {author} {\bibfnamefont {C.~H.~W.}\
  \bibnamefont {{Barnes}}},\ }\href {https://arxiv.org/abs/2005.14475}
  {\bibfield  {journal} {\bibinfo  {journal} {arXiv:2005.14475}\ } (\bibinfo
  {year} {2020})}\BibitemShut {NoStop}%
\bibitem [{\citenamefont {Kraus}(2011)}]{kraus_compressed_2011}%
  \BibitemOpen
  \bibfield  {author} {\bibinfo {author} {\bibfnamefont {B.}~\bibnamefont
  {Kraus}},\ }\href {\doibase 10.1103/PhysRevLett.107.250503} {\bibfield
  {journal} {\bibinfo  {journal} {Phys. Rev. Lett.}\ }\textbf {\bibinfo
  {volume} {107}},\ \bibinfo {pages} {250503} (\bibinfo {year}
  {2011})}\BibitemShut {NoStop}%
\bibitem [{\citenamefont {Hebenstreit}\ \emph {et~al.}(2017)\citenamefont
  {Hebenstreit}, \citenamefont {Alsina}, \citenamefont {Latorre},\ and\
  \citenamefont {Kraus}}]{hebenstreit_compressed_2017}%
  \BibitemOpen
  \bibfield  {author} {\bibinfo {author} {\bibfnamefont {M.}~\bibnamefont
  {Hebenstreit}}, \bibinfo {author} {\bibfnamefont {D.}~\bibnamefont {Alsina}},
  \bibinfo {author} {\bibfnamefont {J.~I.}\ \bibnamefont {Latorre}}, \ and\
  \bibinfo {author} {\bibfnamefont {B.}~\bibnamefont {Kraus}},\ }\href
  {\doibase 10.1103/PhysRevA.95.052339} {\bibfield  {journal} {\bibinfo
  {journal} {Phys. Rev. A}\ }\textbf {\bibinfo {volume} {95}},\ \bibinfo
  {pages} {052339} (\bibinfo {year} {2017})}\BibitemShut {NoStop}%
\bibitem [{\citenamefont {Schmoll}\ and\ \citenamefont
  {Orus}(2017)}]{schmoll_kitaev_2017}%
  \BibitemOpen
  \bibfield  {author} {\bibinfo {author} {\bibfnamefont {P.}~\bibnamefont
  {Schmoll}}\ and\ \bibinfo {author} {\bibfnamefont {R.}~\bibnamefont {Orus}},\
  }\href {\doibase 10.1103/PhysRevB.95.045112} {\bibfield  {journal} {\bibinfo
  {journal} {Phys. Rev. B}\ }\textbf {\bibinfo {volume} {95}},\ \bibinfo
  {pages} {045112} (\bibinfo {year} {2017})}\BibitemShut {NoStop}%
\bibitem [{\citenamefont {Verstraete}\ \emph {et~al.}(2009)\citenamefont
  {Verstraete}, \citenamefont {Cirac},\ and\ \citenamefont
  {Latorre}}]{verstraete_quantum_2009}%
  \BibitemOpen
  \bibfield  {author} {\bibinfo {author} {\bibfnamefont {F.}~\bibnamefont
  {Verstraete}}, \bibinfo {author} {\bibfnamefont {J.~I.}\ \bibnamefont
  {Cirac}}, \ and\ \bibinfo {author} {\bibfnamefont {J.~I.}\ \bibnamefont
  {Latorre}},\ }\href {\doibase 10.1103/PhysRevA.79.032316} {\bibfield
  {journal} {\bibinfo  {journal} {Phys. Rev. A}\ }\textbf {\bibinfo {volume}
  {79}},\ \bibinfo {pages} {032316} (\bibinfo {year} {2009})}\BibitemShut
  {NoStop}%
\bibitem [{\citenamefont {Cervera-Lierta}(2018)}]{cervera-lierta_exact_2018}%
  \BibitemOpen
  \bibfield  {author} {\bibinfo {author} {\bibfnamefont {A.}~\bibnamefont
  {Cervera-Lierta}},\ }\href {\doibase 10.22331/q-2018-12-21-114} {\bibfield
  {journal} {\bibinfo  {journal} {Quantum}\ }\textbf {\bibinfo {volume} {2}},\
  \bibinfo {pages} {114} (\bibinfo {year} {2018})}\BibitemShut {NoStop}%
\bibitem [{\citenamefont {{Montanaro}}\ and\ \citenamefont
  {{Stanisic}}(2020)}]{montanaro_compressed_2020}%
  \BibitemOpen
  \bibfield  {author} {\bibinfo {author} {\bibfnamefont {A.}~\bibnamefont
  {{Montanaro}}}\ and\ \bibinfo {author} {\bibfnamefont {S.}~\bibnamefont
  {{Stanisic}}},\ }\href {https://arxiv.org/abs/2006.01179} {\bibfield
  {journal} {\bibinfo  {journal} {arXiv:2006.01179}\ } (\bibinfo {year}
  {2020})}\BibitemShut {NoStop}%
\bibitem [{\citenamefont {{Farhi}}\ \emph {et~al.}(2014)\citenamefont
  {{Farhi}}, \citenamefont {{Goldstone}},\ and\ \citenamefont
  {{Gutmann}}}]{QAOA}%
  \BibitemOpen
  \bibfield  {author} {\bibinfo {author} {\bibfnamefont {E.}~\bibnamefont
  {{Farhi}}}, \bibinfo {author} {\bibfnamefont {J.}~\bibnamefont
  {{Goldstone}}}, \ and\ \bibinfo {author} {\bibfnamefont {S.}~\bibnamefont
  {{Gutmann}}},\ }\href {http://arxiv.org/abs/1411.4028} {\bibfield  {journal}
  {\bibinfo  {journal} {arXiv:1411.4028}\ } (\bibinfo {year}
  {2014})}\BibitemShut {NoStop}%
\bibitem [{\citenamefont {Lee}\ \emph {et~al.}(2018)\citenamefont {Lee},
  \citenamefont {Huggins}, \citenamefont {Head-Gordon},\ and\ \citenamefont
  {Whaley}}]{lee2018generalized}%
  \BibitemOpen
  \bibfield  {author} {\bibinfo {author} {\bibfnamefont {J.}~\bibnamefont
  {Lee}}, \bibinfo {author} {\bibfnamefont {W.~J.}\ \bibnamefont {Huggins}},
  \bibinfo {author} {\bibfnamefont {M.}~\bibnamefont {Head-Gordon}}, \ and\
  \bibinfo {author} {\bibfnamefont {K.~B.}\ \bibnamefont {Whaley}},\ }\href
  {https://pubs.acs.org/doi/10.1021/acs.jctc.8b01004} {\bibfield  {journal}
  {\bibinfo  {journal} {J. Chem. Theory Comput.}\ }\textbf {\bibinfo {volume}
  {15}},\ \bibinfo {pages} {311} (\bibinfo {year} {2018})}\BibitemShut
  {NoStop}%
\bibitem [{\citenamefont {{Nakaji}}\ and\ \citenamefont
  {{Yamamoto}}(2020)}]{nakaji_expressibility_2020}%
  \BibitemOpen
  \bibfield  {author} {\bibinfo {author} {\bibfnamefont {K.}~\bibnamefont
  {{Nakaji}}}\ and\ \bibinfo {author} {\bibfnamefont {N.}~\bibnamefont
  {{Yamamoto}}},\ }\href {https://arxiv.org/abs/2005.12537} {\bibfield
  {journal} {\bibinfo  {journal} {arXiv:2005.12537}\ } (\bibinfo {year}
  {2020})}\BibitemShut {NoStop}%
\bibitem [{\citenamefont {McClean}\ \emph {et~al.}(2018)\citenamefont
  {McClean}, \citenamefont {Boixo}, \citenamefont {Smelyanskiy}, \citenamefont
  {Babbush},\ and\ \citenamefont {Neven}}]{mcclean_barren_2018}%
  \BibitemOpen
  \bibfield  {author} {\bibinfo {author} {\bibfnamefont {J.~R.}\ \bibnamefont
  {McClean}}, \bibinfo {author} {\bibfnamefont {S.}~\bibnamefont {Boixo}},
  \bibinfo {author} {\bibfnamefont {V.~N.}\ \bibnamefont {Smelyanskiy}},
  \bibinfo {author} {\bibfnamefont {R.}~\bibnamefont {Babbush}}, \ and\
  \bibinfo {author} {\bibfnamefont {H.}~\bibnamefont {Neven}},\ }\href
  {\doibase 10.1038/s41467-018-07090-4} {\bibfield  {journal} {\bibinfo
  {journal} {Nat. Commun.}\ }\textbf {\bibinfo {volume} {9}},\ \bibinfo {pages}
  {4812} (\bibinfo {year} {2018})}\BibitemShut {NoStop}%
\bibitem [{\citenamefont {Mitarai}\ \emph {et~al.}(2019)\citenamefont
  {Mitarai}, \citenamefont {Yan},\ and\ \citenamefont {Fujii}}]{genVQE}%
  \BibitemOpen
  \bibfield  {author} {\bibinfo {author} {\bibfnamefont {K.}~\bibnamefont
  {Mitarai}}, \bibinfo {author} {\bibfnamefont {T.}~\bibnamefont {Yan}}, \ and\
  \bibinfo {author} {\bibfnamefont {K.}~\bibnamefont {Fujii}},\ }\href
  {\doibase 10.1103/PhysRevApplied.11.044087} {\bibfield  {journal} {\bibinfo
  {journal} {Phys. Rev. Appl.}\ }\textbf {\bibinfo {volume} {11}},\ \bibinfo
  {pages} {044087} (\bibinfo {year} {2019})}\BibitemShut {NoStop}%
\bibitem [{\citenamefont {{Cao}}\ \emph {et~al.}(2017)\citenamefont {{Cao}},
  \citenamefont {{Giacomo Guerreschi}},\ and\ \citenamefont
  {{Aspuru-Guzik}}}]{Cao2017}%
  \BibitemOpen
  \bibfield  {author} {\bibinfo {author} {\bibfnamefont {Y.}~\bibnamefont
  {{Cao}}}, \bibinfo {author} {\bibfnamefont {G.}~\bibnamefont {{Giacomo
  Guerreschi}}}, \ and\ \bibinfo {author} {\bibfnamefont {A.}~\bibnamefont
  {{Aspuru-Guzik}}},\ }\href {http://arxiv.org/abs/1711.11240} {\bibfield
  {journal} {\bibinfo  {journal} {arXiv:1711.11240}\ } (\bibinfo {year}
  {2017})}\BibitemShut {NoStop}%
\bibitem [{\citenamefont {Romero}\ \emph {et~al.}(2017)\citenamefont {Romero},
  \citenamefont {Olson},\ and\ \citenamefont {Aspuru-Guzik}}]{Romero2017}%
  \BibitemOpen
  \bibfield  {author} {\bibinfo {author} {\bibfnamefont {J.}~\bibnamefont
  {Romero}}, \bibinfo {author} {\bibfnamefont {J.~P.}\ \bibnamefont {Olson}}, \
  and\ \bibinfo {author} {\bibfnamefont {A.}~\bibnamefont {Aspuru-Guzik}},\
  }\href {\doibase 10.1088/2058-9565/aa8072} {\bibfield  {journal} {\bibinfo
  {journal} {Quantum Sci. Technol.}\ }\textbf {\bibinfo {volume} {2}},\
  \bibinfo {pages} {045001} (\bibinfo {year} {2017})}\BibitemShut {NoStop}%
\bibitem [{\citenamefont {Schuld}\ and\ \citenamefont
  {Petruccione}(2018)}]{schuld_quantum_2018}%
  \BibitemOpen
  \bibfield  {author} {\bibinfo {author} {\bibfnamefont {M.}~\bibnamefont
  {Schuld}}\ and\ \bibinfo {author} {\bibfnamefont {F.}~\bibnamefont
  {Petruccione}},\ }\href {\doibase 10.1038/s41598-018-20403-3} {\bibfield
  {journal} {\bibinfo  {journal} {Sci Rep}\ }\textbf {\bibinfo {volume} {8}},\
  \bibinfo {pages} {2772} (\bibinfo {year} {2018})}\BibitemShut {NoStop}%
\bibitem [{\citenamefont {{Verdon}}\ \emph {et~al.}(2019)\citenamefont
  {{Verdon}}, \citenamefont {{Broughton}}, \citenamefont {{McClean}},
  \citenamefont {{Sung}}, \citenamefont {{Babbush}}, \citenamefont {{Jiang}},
  \citenamefont {{Neven}},\ and\ \citenamefont
  {{Mohseni}}}]{verdon_learning_2019}%
  \BibitemOpen
  \bibfield  {author} {\bibinfo {author} {\bibfnamefont {G.}~\bibnamefont
  {{Verdon}}}, \bibinfo {author} {\bibfnamefont {M.}~\bibnamefont
  {{Broughton}}}, \bibinfo {author} {\bibfnamefont {J.~R.}\ \bibnamefont
  {{McClean}}}, \bibinfo {author} {\bibfnamefont {K.~J.}\ \bibnamefont
  {{Sung}}}, \bibinfo {author} {\bibfnamefont {R.}~\bibnamefont {{Babbush}}},
  \bibinfo {author} {\bibfnamefont {Z.}~\bibnamefont {{Jiang}}}, \bibinfo
  {author} {\bibfnamefont {H.}~\bibnamefont {{Neven}}}, \ and\ \bibinfo
  {author} {\bibfnamefont {M.}~\bibnamefont {{Mohseni}}},\ }\href
  {http://arxiv.org/abs/1907.05415} {\bibfield  {journal} {\bibinfo  {journal}
  {arXiv:1907.05415}\ } (\bibinfo {year} {2019})}\BibitemShut {NoStop}%
\bibitem [{\citenamefont {Schuld}\ \emph {et~al.}(2020)\citenamefont {Schuld},
  \citenamefont {Bocharov}, \citenamefont {Svore},\ and\ \citenamefont
  {Wiebe}}]{schuld_circuit-centric_2020}%
  \BibitemOpen
  \bibfield  {author} {\bibinfo {author} {\bibfnamefont {M.}~\bibnamefont
  {Schuld}}, \bibinfo {author} {\bibfnamefont {A.}~\bibnamefont {Bocharov}},
  \bibinfo {author} {\bibfnamefont {K.}~\bibnamefont {Svore}}, \ and\ \bibinfo
  {author} {\bibfnamefont {N.}~\bibnamefont {Wiebe}},\ }\href {\doibase
  10.1103/PhysRevA.101.032308} {\bibfield  {journal} {\bibinfo  {journal}
  {Phys. Rev. A}\ }\textbf {\bibinfo {volume} {101}},\ \bibinfo {pages}
  {032308} (\bibinfo {year} {2020})}\BibitemShut {NoStop}%
\bibitem [{\citenamefont {Pérez-Salinas}\ \emph {et~al.}(2020)\citenamefont
  {Pérez-Salinas}, \citenamefont {Cervera-Lierta}, \citenamefont
  {Gil-Fuster},\ and\ \citenamefont {Latorre}}]{perez-salinas_data_2020}%
  \BibitemOpen
  \bibfield  {author} {\bibinfo {author} {\bibfnamefont {A.}~\bibnamefont
  {Pérez-Salinas}}, \bibinfo {author} {\bibfnamefont {A.}~\bibnamefont
  {Cervera-Lierta}}, \bibinfo {author} {\bibfnamefont {E.}~\bibnamefont
  {Gil-Fuster}}, \ and\ \bibinfo {author} {\bibfnamefont {J.~I.}\ \bibnamefont
  {Latorre}},\ }\href {\doibase 10.22331/q-2020-02-06-226} {\bibfield
  {journal} {\bibinfo  {journal} {Quantum}\ }\textbf {\bibinfo {volume} {4}},\
  \bibinfo {pages} {226} (\bibinfo {year} {2020})}\BibitemShut {NoStop}%
\bibitem [{\citenamefont {{Abbas}}\ \emph {et~al.}(2020)\citenamefont
  {{Abbas}}, \citenamefont {{Schuld}},\ and\ \citenamefont
  {{Petruccione}}}]{abbas_quantum_2020}%
  \BibitemOpen
  \bibfield  {author} {\bibinfo {author} {\bibfnamefont {A.}~\bibnamefont
  {{Abbas}}}, \bibinfo {author} {\bibfnamefont {M.}~\bibnamefont {{Schuld}}}, \
  and\ \bibinfo {author} {\bibfnamefont {F.}~\bibnamefont {{Petruccione}}},\
  }\href {http://arxiv.org/abs/2001.10833} {\bibfield  {journal} {\bibinfo
  {journal} {arXiv:2001.10833}\ } (\bibinfo {year} {2020})}\BibitemShut
  {NoStop}%
\bibitem [{\citenamefont {Zhang}\ and\ \citenamefont {Yin}(2020)}]{cVQE}%
  \BibitemOpen
  \bibfield  {author} {\bibinfo {author} {\bibfnamefont {D.-B.}\ \bibnamefont
  {Zhang}}\ and\ \bibinfo {author} {\bibfnamefont {T.}~\bibnamefont {Yin}},\
  }\href {\doibase 10.1103/PhysRevA.101.032311} {\bibfield  {journal} {\bibinfo
   {journal} {Phys. Rev. A}\ }\textbf {\bibinfo {volume} {101}},\ \bibinfo
  {pages} {032311} (\bibinfo {year} {2020})}\BibitemShut {NoStop}%
\bibitem [{\citenamefont {{Kyaw}}\ \emph {et~al.}(2020)\citenamefont {{Kyaw}},
  \citenamefont {{Menke}}, \citenamefont {{Sim}}, \citenamefont {{Sawaya}},
  \citenamefont {{Oliver}}, \citenamefont {{Giacomo Guerreschi}},\ and\
  \citenamefont {{Aspuru-Guzik}}}]{QCAD}%
  \BibitemOpen
  \bibfield  {author} {\bibinfo {author} {\bibfnamefont {T.~H.}\ \bibnamefont
  {{Kyaw}}}, \bibinfo {author} {\bibfnamefont {T.}~\bibnamefont {{Menke}}},
  \bibinfo {author} {\bibfnamefont {S.}~\bibnamefont {{Sim}}}, \bibinfo
  {author} {\bibfnamefont {N.~P.~D.}\ \bibnamefont {{Sawaya}}}, \bibinfo
  {author} {\bibfnamefont {W.~D.}\ \bibnamefont {{Oliver}}}, \bibinfo {author}
  {\bibfnamefont {G.}~\bibnamefont {{Giacomo Guerreschi}}}, \ and\ \bibinfo
  {author} {\bibfnamefont {A.}~\bibnamefont {{Aspuru-Guzik}}},\ }\href
  {http://arxiv.org/abs/2006.03070} {\bibfield  {journal} {\bibinfo  {journal}
  {arXiv:2006.03070}\ } (\bibinfo {year} {2020})}\BibitemShut {NoStop}%
\bibitem [{\citenamefont {Kottmann}\ \emph {et~al.}(2020)\citenamefont
  {Kottmann}, \citenamefont {Alperin-Lea}, \citenamefont {Tamayo-Mendoza},
  \citenamefont {Cervera-Lierta}, \citenamefont {Lavigne}, \citenamefont {Yen},
  \citenamefont {Verteletskyi}, \citenamefont {Schleich}, \citenamefont
  {Anand}, \citenamefont {Degroote} \emph {et~al.}}]{tequila}%
  \BibitemOpen
  \bibfield  {author} {\bibinfo {author} {\bibfnamefont {J.~S.}\ \bibnamefont
  {Kottmann}}, \bibinfo {author} {\bibfnamefont {S.}~\bibnamefont
  {Alperin-Lea}}, \bibinfo {author} {\bibfnamefont {T.}~\bibnamefont
  {Tamayo-Mendoza}}, \bibinfo {author} {\bibfnamefont {A.}~\bibnamefont
  {Cervera-Lierta}}, \bibinfo {author} {\bibfnamefont {C.}~\bibnamefont
  {Lavigne}}, \bibinfo {author} {\bibfnamefont {T.-C.}\ \bibnamefont {Yen}},
  \bibinfo {author} {\bibfnamefont {V.}~\bibnamefont {Verteletskyi}}, \bibinfo
  {author} {\bibfnamefont {P.}~\bibnamefont {Schleich}}, \bibinfo {author}
  {\bibfnamefont {A.}~\bibnamefont {Anand}}, \bibinfo {author} {\bibfnamefont
  {M.}~\bibnamefont {Degroote}},  \emph {et~al.},\ }\href
  {https://arxiv.org/abs/2011.03057} {\bibfield  {journal} {\bibinfo  {journal}
  {arXiv:2011.03057}\ } (\bibinfo {year} {2020})}\BibitemShut {NoStop}%
\bibitem [{\citenamefont {Suzuki}\ \emph {et~al.}(2020)\citenamefont {Suzuki},
  \citenamefont {Kawase}, \citenamefont {Masumura}, \citenamefont {Hiraga},
  \citenamefont {Nakadai}, \citenamefont {Chen}, \citenamefont {Nakanishi},
  \citenamefont {Mitarai}, \citenamefont {Imai}, \citenamefont {Tamiya} \emph
  {et~al.}}]{qulacs}%
  \BibitemOpen
  \bibfield  {author} {\bibinfo {author} {\bibfnamefont {Y.}~\bibnamefont
  {Suzuki}}, \bibinfo {author} {\bibfnamefont {Y.}~\bibnamefont {Kawase}},
  \bibinfo {author} {\bibfnamefont {Y.}~\bibnamefont {Masumura}}, \bibinfo
  {author} {\bibfnamefont {Y.}~\bibnamefont {Hiraga}}, \bibinfo {author}
  {\bibfnamefont {M.}~\bibnamefont {Nakadai}}, \bibinfo {author} {\bibfnamefont
  {J.}~\bibnamefont {Chen}}, \bibinfo {author} {\bibfnamefont {K.~M.}\
  \bibnamefont {Nakanishi}}, \bibinfo {author} {\bibfnamefont {K.}~\bibnamefont
  {Mitarai}}, \bibinfo {author} {\bibfnamefont {R.}~\bibnamefont {Imai}},
  \bibinfo {author} {\bibfnamefont {S.}~\bibnamefont {Tamiya}},  \emph
  {et~al.},\ }\href {https://arxiv.org/abs/2011.13524} {\bibfield  {journal}
  {\bibinfo  {journal} {arXiv:2011.13524}\ } (\bibinfo {year}
  {2020})}\BibitemShut {NoStop}%
\bibitem [{\citenamefont {Langari}(1998)}]{langari_phase_1998}%
  \BibitemOpen
  \bibfield  {author} {\bibinfo {author} {\bibfnamefont {A.}~\bibnamefont
  {Langari}},\ }\href {\doibase 10.1103/PhysRevB.58.14467} {\bibfield
  {journal} {\bibinfo  {journal} {Phys. Rev. B}\ }\textbf {\bibinfo {volume}
  {58}},\ \bibinfo {pages} {14467} (\bibinfo {year} {1998})}\BibitemShut
  {NoStop}%
\bibitem [{\citenamefont {Van~Dyke}\ \emph {et~al.}(2021)\citenamefont
  {Van~Dyke}, \citenamefont {Barron}, \citenamefont {Mayhall}, \citenamefont
  {Barnes},\ and\ \citenamefont {Economou}}]{van2021preparing}%
  \BibitemOpen
  \bibfield  {author} {\bibinfo {author} {\bibfnamefont {J.~S.}\ \bibnamefont
  {Van~Dyke}}, \bibinfo {author} {\bibfnamefont {G.~S.}\ \bibnamefont
  {Barron}}, \bibinfo {author} {\bibfnamefont {N.~J.}\ \bibnamefont {Mayhall}},
  \bibinfo {author} {\bibfnamefont {E.}~\bibnamefont {Barnes}}, \ and\ \bibinfo
  {author} {\bibfnamefont {S.~E.}\ \bibnamefont {Economou}},\ }\href
  {https://arxiv.org/abs/2103.13388} {\bibfield  {journal} {\bibinfo  {journal}
  {arXiv:2103.13388}\ } (\bibinfo {year} {2021})}\BibitemShut {NoStop}%
\bibitem [{\citenamefont {Cirac}\ and\ \citenamefont {Sierra}(2010)}]{HS-XXZ}%
  \BibitemOpen
  \bibfield  {author} {\bibinfo {author} {\bibfnamefont {J.~I.}\ \bibnamefont
  {Cirac}}\ and\ \bibinfo {author} {\bibfnamefont {G.}~\bibnamefont {Sierra}},\
  }\href {\doibase 10.1103/PhysRevB.81.104431} {\bibfield  {journal} {\bibinfo
  {journal} {Phys. Rev. B}\ }\textbf {\bibinfo {volume} {81}},\ \bibinfo
  {pages} {104431} (\bibinfo {year} {2010})}\BibitemShut {NoStop}%
\bibitem [{\citenamefont {McArdle}\ \emph {et~al.}(2020)\citenamefont
  {McArdle}, \citenamefont {Endo}, \citenamefont {Aspuru-Guzik}, \citenamefont
  {Benjamin},\ and\ \citenamefont {Yuan}}]{mcardle2020quantumreview}%
  \BibitemOpen
  \bibfield  {author} {\bibinfo {author} {\bibfnamefont {S.}~\bibnamefont
  {McArdle}}, \bibinfo {author} {\bibfnamefont {S.}~\bibnamefont {Endo}},
  \bibinfo {author} {\bibfnamefont {A.}~\bibnamefont {Aspuru-Guzik}}, \bibinfo
  {author} {\bibfnamefont {S.~C.}\ \bibnamefont {Benjamin}}, \ and\ \bibinfo
  {author} {\bibfnamefont {X.}~\bibnamefont {Yuan}},\ }\href
  {https://journals.aps.org/rmp/abstract/10.1103/RevModPhys.92.015003}
  {\bibfield  {journal} {\bibinfo  {journal} {Rev. Mod. Phys.}\ }\textbf
  {\bibinfo {volume} {92}},\ \bibinfo {pages} {015003} (\bibinfo {year}
  {2020})}\BibitemShut {NoStop}%
\bibitem [{\citenamefont {Kottmann}\ \emph {et~al.}(2021)\citenamefont
  {Kottmann}, \citenamefont {Anand},\ and\ \citenamefont
  {Aspuru-Guzik}}]{kottmann2021feasible}%
  \BibitemOpen
  \bibfield  {author} {\bibinfo {author} {\bibfnamefont {J.~S.}\ \bibnamefont
  {Kottmann}}, \bibinfo {author} {\bibfnamefont {A.}~\bibnamefont {Anand}}, \
  and\ \bibinfo {author} {\bibfnamefont {A.}~\bibnamefont {Aspuru-Guzik}},\
  }\href {\doibase 10.1039/D0SC06627C} {\bibfield  {journal} {\bibinfo
  {journal} {Chem. Sci.}\ ,\ } (\bibinfo {year} {2021})}\BibitemShut {NoStop}%
\bibitem [{\citenamefont {{Kottmann}}\ \emph {et~al.}(2020)\citenamefont
  {{Kottmann}}, \citenamefont {{Schleich}}, \citenamefont {{Tamayo-Mendoza}},\
  and\ \citenamefont {{Aspuru-Guzik}}}]{kottmann2020reducing}%
  \BibitemOpen
  \bibfield  {author} {\bibinfo {author} {\bibfnamefont {J.~S.}\ \bibnamefont
  {{Kottmann}}}, \bibinfo {author} {\bibfnamefont {P.}~\bibnamefont
  {{Schleich}}}, \bibinfo {author} {\bibfnamefont {T.}~\bibnamefont
  {{Tamayo-Mendoza}}}, \ and\ \bibinfo {author} {\bibfnamefont
  {A.}~\bibnamefont {{Aspuru-Guzik}}},\ }\href
  {http://arxiv.org/abs/2008.02819} {\bibfield  {journal} {\bibinfo  {journal}
  {arXiv:2008.02819}\ } (\bibinfo {year} {2020})}\BibitemShut {NoStop}%
\bibitem [{\citenamefont {Sawaya}\ \emph {et~al.}(2020)\citenamefont {Sawaya},
  \citenamefont {Menke}, \citenamefont {Kyaw}, \citenamefont {Johri},
  \citenamefont {Aspuru-Guzik},\ and\ \citenamefont {Guerreschi}}]{d-level}%
  \BibitemOpen
  \bibfield  {author} {\bibinfo {author} {\bibfnamefont {N.~P.~D.}\
  \bibnamefont {Sawaya}}, \bibinfo {author} {\bibfnamefont {T.}~\bibnamefont
  {Menke}}, \bibinfo {author} {\bibfnamefont {T.~H.}\ \bibnamefont {Kyaw}},
  \bibinfo {author} {\bibfnamefont {S.}~\bibnamefont {Johri}}, \bibinfo
  {author} {\bibfnamefont {A.}~\bibnamefont {Aspuru-Guzik}}, \ and\ \bibinfo
  {author} {\bibfnamefont {G.~G.}\ \bibnamefont {Guerreschi}},\ }\href
  {https://doi.org/10.1038/s41534-020-0278-0} {\bibfield  {journal} {\bibinfo
  {journal} {NPJ Quantum Inf.}\ }\textbf {\bibinfo {volume} {6}} (\bibinfo
  {year} {2020})}\BibitemShut {NoStop}%
\bibitem [{\citenamefont {Skolik}\ \emph {et~al.}(2020)\citenamefont {Skolik},
  \citenamefont {McClean}, \citenamefont {Mohseni}, \citenamefont {van~der
  Smagt},\ and\ \citenamefont {Leib}}]{lw_VQE}%
  \BibitemOpen
  \bibfield  {author} {\bibinfo {author} {\bibfnamefont {A.}~\bibnamefont
  {Skolik}}, \bibinfo {author} {\bibfnamefont {J.~R.}\ \bibnamefont {McClean}},
  \bibinfo {author} {\bibfnamefont {M.}~\bibnamefont {Mohseni}}, \bibinfo
  {author} {\bibfnamefont {P.}~\bibnamefont {van~der Smagt}}, \ and\ \bibinfo
  {author} {\bibfnamefont {M.}~\bibnamefont {Leib}},\ }\href
  {https://arxiv.org/abs/2006.14904} {\bibfield  {journal} {\bibinfo  {journal}
  {arXiv:2006.14904}\ } (\bibinfo {year} {2020})}\BibitemShut {NoStop}%
\bibitem [{\citenamefont {Cerezo}\ \emph {et~al.}(2020)\citenamefont {Cerezo},
  \citenamefont {Sone}, \citenamefont {Volkoff}, \citenamefont {Cincio},\ and\
  \citenamefont {Coles}}]{cerezo_cost-function-dependent_2020}%
  \BibitemOpen
  \bibfield  {author} {\bibinfo {author} {\bibfnamefont {M.}~\bibnamefont
  {Cerezo}}, \bibinfo {author} {\bibfnamefont {A.}~\bibnamefont {Sone}},
  \bibinfo {author} {\bibfnamefont {T.}~\bibnamefont {Volkoff}}, \bibinfo
  {author} {\bibfnamefont {L.}~\bibnamefont {Cincio}}, \ and\ \bibinfo {author}
  {\bibfnamefont {P.~J.}\ \bibnamefont {Coles}},\ }\href
  {http://arxiv.org/abs/2001.00550} {\bibfield  {journal} {\bibinfo  {journal}
  {arXiv:2001.00550 [quant-ph]}\ } (\bibinfo {year} {2020})}\BibitemShut
  {NoStop}%
\bibitem [{\citenamefont {{Premaratne}}\ and\ \citenamefont
  {{Matsuura}}(2020)}]{premaratne_engineering_2020}%
  \BibitemOpen
  \bibfield  {author} {\bibinfo {author} {\bibfnamefont {S.~P.}\ \bibnamefont
  {{Premaratne}}}\ and\ \bibinfo {author} {\bibfnamefont {A.~Y.}\ \bibnamefont
  {{Matsuura}}},\ }\href {https://arxiv.org/abs/2006.03747} {\bibfield
  {journal} {\bibinfo  {journal} {arXiv:2006.03747}\ } (\bibinfo {year}
  {2020})}\BibitemShut {NoStop}%
\bibitem [{\citenamefont {{Volkoff}}\ and\ \citenamefont
  {{Coles}}(2020)}]{volkoff_large_2020}%
  \BibitemOpen
  \bibfield  {author} {\bibinfo {author} {\bibfnamefont {T.}~\bibnamefont
  {{Volkoff}}}\ and\ \bibinfo {author} {\bibfnamefont {P.~J.}\ \bibnamefont
  {{Coles}}},\ }\href {http://arxiv.org/abs/2005.12200} {\bibfield  {journal}
  {\bibinfo  {journal} {arXiv:2005.12200}\ } (\bibinfo {year}
  {2020})}\BibitemShut {NoStop}%
\bibitem [{\citenamefont {Grant}\ \emph {et~al.}(2019)\citenamefont {Grant},
  \citenamefont {Wossnig}, \citenamefont {Ostaszewski},\ and\ \citenamefont
  {Benedetti}}]{grant_initialization_2019}%
  \BibitemOpen
  \bibfield  {author} {\bibinfo {author} {\bibfnamefont {E.}~\bibnamefont
  {Grant}}, \bibinfo {author} {\bibfnamefont {L.}~\bibnamefont {Wossnig}},
  \bibinfo {author} {\bibfnamefont {M.}~\bibnamefont {Ostaszewski}}, \ and\
  \bibinfo {author} {\bibfnamefont {M.}~\bibnamefont {Benedetti}},\ }\href
  {\doibase 10.22331/q-2019-12-09-214} {\bibfield  {journal} {\bibinfo
  {journal} {Quantum}\ }\textbf {\bibinfo {volume} {3}},\ \bibinfo {pages}
  {214} (\bibinfo {year} {2019})}\BibitemShut {NoStop}%
\bibitem [{\citenamefont {{Skolik}}\ \emph {et~al.}(2020)\citenamefont
  {{Skolik}}, \citenamefont {{McClean}}, \citenamefont {{Mohseni}},
  \citenamefont {{van der Smagt}},\ and\ \citenamefont
  {{Leib}}}]{skolik_layerwise_2020}%
  \BibitemOpen
  \bibfield  {author} {\bibinfo {author} {\bibfnamefont {A.}~\bibnamefont
  {{Skolik}}}, \bibinfo {author} {\bibfnamefont {J.~R.}\ \bibnamefont
  {{McClean}}}, \bibinfo {author} {\bibfnamefont {M.}~\bibnamefont
  {{Mohseni}}}, \bibinfo {author} {\bibfnamefont {P.}~\bibnamefont {{van der
  Smagt}}}, \ and\ \bibinfo {author} {\bibfnamefont {M.}~\bibnamefont
  {{Leib}}},\ }\href {http://arxiv.org/abs/2006.14904} {\bibfield  {journal}
  {\bibinfo  {journal} {arXiv:2006.14904}\ } (\bibinfo {year}
  {2020})}\BibitemShut {NoStop}%
\bibitem [{\citenamefont {Wang}\ \emph {et~al.}(2020)\citenamefont {Wang},
  \citenamefont {Fontana}, \citenamefont {Cerezo}, \citenamefont {Sharma},
  \citenamefont {Sone}, \citenamefont {Cincio},\ and\ \citenamefont
  {Coles}}]{wang_noise-induced_2020}%
  \BibitemOpen
  \bibfield  {author} {\bibinfo {author} {\bibfnamefont {S.}~\bibnamefont
  {Wang}}, \bibinfo {author} {\bibfnamefont {E.}~\bibnamefont {Fontana}},
  \bibinfo {author} {\bibfnamefont {M.}~\bibnamefont {Cerezo}}, \bibinfo
  {author} {\bibfnamefont {K.}~\bibnamefont {Sharma}}, \bibinfo {author}
  {\bibfnamefont {A.}~\bibnamefont {Sone}}, \bibinfo {author} {\bibfnamefont
  {L.}~\bibnamefont {Cincio}}, \ and\ \bibinfo {author} {\bibfnamefont {P.~J.}\
  \bibnamefont {Coles}},\ }\href {http://arxiv.org/abs/2007.14384} {\bibfield
  {journal} {\bibinfo  {journal} {arXiv:2007.14384}\ } (\bibinfo {year}
  {2020})}\BibitemShut {NoStop}%
\bibitem [{\citenamefont {Barkoutsos}\ \emph {et~al.}(2018)\citenamefont
  {Barkoutsos}, \citenamefont {Gonthier}, \citenamefont {Sokolov},
  \citenamefont {Moll}, \citenamefont {Salis}, \citenamefont {Fuhrer},
  \citenamefont {Ganzhorn}, \citenamefont {Egger}, \citenamefont {Troyer},
  \citenamefont {Mezzacapo}, \citenamefont {Filipp},\ and\ \citenamefont
  {Tavernelli}}]{barkoutsos_quantum_2018}%
  \BibitemOpen
  \bibfield  {author} {\bibinfo {author} {\bibfnamefont {P.~K.}\ \bibnamefont
  {Barkoutsos}}, \bibinfo {author} {\bibfnamefont {J.~F.}\ \bibnamefont
  {Gonthier}}, \bibinfo {author} {\bibfnamefont {I.}~\bibnamefont {Sokolov}},
  \bibinfo {author} {\bibfnamefont {N.}~\bibnamefont {Moll}}, \bibinfo {author}
  {\bibfnamefont {G.}~\bibnamefont {Salis}}, \bibinfo {author} {\bibfnamefont
  {A.}~\bibnamefont {Fuhrer}}, \bibinfo {author} {\bibfnamefont
  {M.}~\bibnamefont {Ganzhorn}}, \bibinfo {author} {\bibfnamefont {D.~J.}\
  \bibnamefont {Egger}}, \bibinfo {author} {\bibfnamefont {M.}~\bibnamefont
  {Troyer}}, \bibinfo {author} {\bibfnamefont {A.}~\bibnamefont {Mezzacapo}},
  \bibinfo {author} {\bibfnamefont {S.}~\bibnamefont {Filipp}}, \ and\ \bibinfo
  {author} {\bibfnamefont {I.}~\bibnamefont {Tavernelli}},\ }\href {\doibase
  10.1103/PhysRevA.98.022322} {\bibfield  {journal} {\bibinfo  {journal} {Phys.
  Rev. A}\ }\textbf {\bibinfo {volume} {98}},\ \bibinfo {pages} {022322}
  (\bibinfo {year} {2018})}\BibitemShut {NoStop}%
\bibitem [{\citenamefont {Cowtan}\ \emph {et~al.}(2020)\citenamefont {Cowtan},
  \citenamefont {Simmons},\ and\ \citenamefont {Duncan}}]{cowtan_generic_2020}%
  \BibitemOpen
  \bibfield  {author} {\bibinfo {author} {\bibfnamefont {A.}~\bibnamefont
  {Cowtan}}, \bibinfo {author} {\bibfnamefont {W.}~\bibnamefont {Simmons}}, \
  and\ \bibinfo {author} {\bibfnamefont {R.}~\bibnamefont {Duncan}},\ }\href
  {https://arxiv.org/abs/2007.10515} {\bibfield  {journal} {\bibinfo  {journal}
  {arXiv:2007.10515}\ } (\bibinfo {year} {2020})}\BibitemShut {NoStop}%
\bibitem [{\citenamefont {Fujii}\ \emph {et~al.}(2020)\citenamefont {Fujii},
  \citenamefont {Mitarai}, \citenamefont {Mizukami},\ and\ \citenamefont
  {Nakagawa}}]{fujii_deep_2020}%
  \BibitemOpen
  \bibfield  {author} {\bibinfo {author} {\bibfnamefont {K.}~\bibnamefont
  {Fujii}}, \bibinfo {author} {\bibfnamefont {K.}~\bibnamefont {Mitarai}},
  \bibinfo {author} {\bibfnamefont {W.}~\bibnamefont {Mizukami}}, \ and\
  \bibinfo {author} {\bibfnamefont {Y.~O.}\ \bibnamefont {Nakagawa}},\ }\href
  {http://arxiv.org/abs/2007.10917} {\bibfield  {journal} {\bibinfo  {journal}
  {arXiv:2007.10917}\ } (\bibinfo {year} {2020})}\BibitemShut {NoStop}%
\bibitem [{\citenamefont {Huggins}\ \emph {et~al.}(2020)\citenamefont
  {Huggins}, \citenamefont {Lee}, \citenamefont {Baek}, \citenamefont
  {O’Gorman},\ and\ \citenamefont {Whaley}}]{huggins_non-orthogonal_2020}%
  \BibitemOpen
  \bibfield  {author} {\bibinfo {author} {\bibfnamefont {W.~J.}\ \bibnamefont
  {Huggins}}, \bibinfo {author} {\bibfnamefont {J.}~\bibnamefont {Lee}},
  \bibinfo {author} {\bibfnamefont {U.}~\bibnamefont {Baek}}, \bibinfo {author}
  {\bibfnamefont {B.}~\bibnamefont {O’Gorman}}, \ and\ \bibinfo {author}
  {\bibfnamefont {K.~B.}\ \bibnamefont {Whaley}},\ }\href {\doibase
  10.1088/1367-2630/ab867b} {\bibfield  {journal} {\bibinfo  {journal} {New J.
  Phys.}\ }\textbf {\bibinfo {volume} {22}},\ \bibinfo {pages} {073009}
  (\bibinfo {year} {2020})}\BibitemShut {NoStop}%
\bibitem [{\citenamefont {{Chivilikhin}}\ \emph {et~al.}(2020)\citenamefont
  {{Chivilikhin}}, \citenamefont {{Samarin}}, \citenamefont {{Ulyantsev}},
  \citenamefont {{Iorsh}}, \citenamefont {{Oganov}},\ and\ \citenamefont
  {{Kyriienko}}}]{chivilikhin_mog-vqe_2020}%
  \BibitemOpen
  \bibfield  {author} {\bibinfo {author} {\bibfnamefont {D.}~\bibnamefont
  {{Chivilikhin}}}, \bibinfo {author} {\bibfnamefont {A.}~\bibnamefont
  {{Samarin}}}, \bibinfo {author} {\bibfnamefont {V.}~\bibnamefont
  {{Ulyantsev}}}, \bibinfo {author} {\bibfnamefont {I.}~\bibnamefont
  {{Iorsh}}}, \bibinfo {author} {\bibfnamefont {A.~R.}\ \bibnamefont
  {{Oganov}}}, \ and\ \bibinfo {author} {\bibfnamefont {O.}~\bibnamefont
  {{Kyriienko}}},\ }\href {https://arxiv.org/abs/2007.04424} {\bibfield
  {journal} {\bibinfo  {journal} {arXiv:2007.04424}\ } (\bibinfo {year}
  {2020})}\BibitemShut {NoStop}%
\bibitem [{\citenamefont {Endo}\ \emph {et~al.}(2018)\citenamefont {Endo},
  \citenamefont {Benjamin},\ and\ \citenamefont {Li}}]{suguru2018practical}%
  \BibitemOpen
  \bibfield  {author} {\bibinfo {author} {\bibfnamefont {S.}~\bibnamefont
  {Endo}}, \bibinfo {author} {\bibfnamefont {S.~C.}\ \bibnamefont {Benjamin}},
  \ and\ \bibinfo {author} {\bibfnamefont {Y.}~\bibnamefont {Li}},\ }\href
  {\doibase 10.1103/PhysRevX.8.031027} {\bibfield  {journal} {\bibinfo
  {journal} {Phys. Rev. X}\ }\textbf {\bibinfo {volume} {8}},\ \bibinfo {pages}
  {031027} (\bibinfo {year} {2018})}\BibitemShut {NoStop}%
\end{thebibliography}%

\end{document}